\documentclass[submission, Codebases]{SciPost}

\usepackage{soul}
\usepackage{color}
\usepackage{floatrow}
\usepackage{amsfonts}
\usepackage{booktabs}
\usepackage{listings}
\usepackage[theme=default]{jlcode}
\usepackage{doi}
\usepackage{hyperref}
\usepackage{subfig}

\newcommand*\diff{\mathop{}\!\mathrm{d}}

\newcommand{\dn}{\text{dn}}
\newcommand{\sn}{\text{sn}}
\newcommand{\cn}{\text{cn}}
\newcommand{\half}{\frac{1}{2}}

\newcounter{magicrownumbers}
\newcommand\rownumber{\stepcounter{magicrownumbers}\arabic{magicrownumbers}}

\definecolor{codegreen}{rgb}{0,0.5,0.6}
\definecolor{codegray}{rgb}{0.5,0.5,0.5}
\definecolor{codepurple}{rgb}{0.58,0,0.82}
\definecolor{backcolour}{rgb}{0.95,0.95,0.95}

\lstdefinestyle{mystyle}{
    backgroundcolor=\color{backcolour},   
    commentstyle=\color{codegreen},
    numberstyle=\tiny\color{codegray},
    stringstyle=\color{codepurple},
    captionpos=b,                    
    numbersep=10pt,                  
}

\definecolor{darkgreen}{RGB}{0,138,16}
\definecolor{darkblue}{RGB}{0,12,176}

\hypersetup{
 colorlinks=true,
 linkcolor=blue,
 filecolor=red,
 citecolor=darkgreen,      
 urlcolor=cyan,
}
 
\hypersetup{linktocpage}

\begin{document}
\floatsetup[figure]{style=plain,subcapbesideposition=top}
\floatsetup[table]{capposition=top}

\begin{center}{\Large \textbf{
NonlinearSchrodinger: Higher-Order Algorithms and Darboux Transformations for Nonlinear Schr\"odinger Equations
}}\end{center}

\begin{center}
Omar A. Ashour\textsuperscript{1}
\end{center}

\begin{center}
{\bf 1} Department of Physics, University of California, Berkeley, CA, 94720
\\
* ashour@berkeley.edu
\end{center}

\begin{center}
\today
\end{center}


\section*{Abstract}
{\bf
\texttt{NonlinearSchrodinger.jl} is a Julia package with a simple interface for studying solutions of nonlinear Schr\"odinger equations (NLSEs). In approximately ten lines of code, one can perform a simulation of the cubic NLSE using one of 32 algorithms, including symplectic and Runge-Kutta-Nystr\"om integrators up to eighth order. Furthermore, it is possible to compute analytical solutions via a numerical implementation of the Darboux transformation for extended NLSEs up to fifth order, with an equally simple interface. In what follows, we review the fundamentals of solving this class of equations numerically and analytically, discuss the implementation, and provide several examples.
}

\vspace{10pt}
\noindent\rule{\textwidth}{1pt}
\tableofcontents\thispagestyle{fancy}
\noindent\rule{\textwidth}{1pt}
\vspace{10pt}

\section{Introduction \label{sec:intro}}
Nonlinear Schr\"odinger equations (NLSEs) are of utmost importance in many fields of physics, including light propagation in nonlinear media \cite{Solli2007, Frisquet2014, Armaroli2015}, Bose-Einstein condensates \cite{Zong2014}, Heisenberg spin chains \cite{Lakshmanan1988, Porsezian1992}, ocean surface waves \cite{Kharif2003, Janssen2003}, and many others. There is a rich body of theoretical studies, guided by computational tools, exploring these equations \cite{Nikolic2017, Kedziora2012a, Kedziora2012, Nikolic2019, Kedziora2011a, Ankiewicz2016, Chin2015, Kedziora2015}.

One central issue with studies in this field is the closed-source nature of the codes used in most works, if not all. While one can find a few packages for solving the cubic nonlinear Schr\"odinger equation on, e.g., GitHub and MATLAB Central, these packages generally have a complicated interface and only implement first or second-order integrators. These integrators are insufficiently accurate for all but the simplest simulations and should not be used in research applications. Moreover, to the best of our knowledge, there are no open-source packages that implement the Darboux transformation for the cubic NLSE, let alone extended NLSEs, as this work does. 

We believe that a modern open-source package, such as the one presented in this work, is necessary for this community. The simple interface streamlines simulations and Darboux transformation calculations, alleviating the need to ``reinvent the wheel'' by every research group. We also expect that it will lower the barrier to entry for new researchers and aid in reproducing results. While we do not introduce any new algorithms or methods in this package, it is designed to perform any simulation or calculation in approximately ten lines of code, no matter how complicated the solution. The uniqueness of \texttt{NonlinearSchrodinger.jl}, coupled with its simple interface and algorithms specialized for the problem at hand, make this work original.

This work is divided as follows: in Sec. \ref{sec:nlse}, we introduce the cubic NLSE and its most notable analytical solutions: the soliton and the breather. We additionally examine its first three integrals of motion. Following this introduction, we provide a brief discussion of solving the NLSE numerically in Sec. \ref{sec:num_soln}. We review splitting methods and higher-order integrators. Furthermore, we examine the numerical implementation, benchmark the algorithms, and provide three examples of using the package for simulations. 

In Sec. \ref{sec:analytical}, we introduce the extended NLSE that our package is equipped to solve via the Darboux transformation. We present the Lax system of the cubic NLSE, the Darboux transformation scheme, and the seed solutions. Furthermore, we provide six examples of using the package for Darboux transformation calculations. Finally, we offer future directions and concluding remarks in Sec. \ref{sec:conc}.

\section{The Nonlinear Schr\"odinger Equation \label{sec:nlse}}
The cubic (1+1) Nonlinear Sch\"odinger equation (NLSE) with anomalous dispersion is given in dimensionless form as
\begin{align}
    i \frac{\partial \psi}{\partial x} + \frac{1}{2} \frac{\partial^2 \psi}{\partial t^2} + |\psi|^2\psi = 0,
    \label{eq:nlse}
\end{align}
where $\psi(x,t)$ is a complex field. The constants in front of each term can be changed by appropriate scaling of $\psi$, $x$ and $t$.

The physical meaning of $x$, $t$, and $\psi$ depend on the problem being studied. For example, in optical fibers, the longitudinal (evolution) variable $x$ represents the distance along the fiber, $t$ is the retarded time (i.e., in the frame moving at the group velocity of the pulse), while $\psi$ is proportional to the slowly varying envelope of the pulse. We chose this convention for the package since it is the norm in many works in this field.
\subsection{Notable Analytical Solutions \label{sec:analytical_soln}}
The NLSE and its extensions can be solved by multiple methods, including the inverse scattering transform \cite{Zakharov1972}, the Hirota bilinear method \cite{Hirota1973}, and the Darboux Transformation \cite{Matveev1991,Kedziora2011a,Kedziora2015,Kedziora2014, Nikolic2017, Nikolic2019}, discussed later in this work. The NLSE has multiple notable solutions, but we highlight only two of them: the soliton and the breather.
\subsubsection{Solitons \label{sec:solitons}}
Solitons are self-focusing wave-packets that maintain their shape due to a balance between nonlinearity and dispersion during propagation. The first-order (fundamental) soliton solution of the NLSE is given by
\begin{align}
    \psi(x,t) = \frac{2 \nu e^{2 i \nu^2 x}}{\cosh(2 \nu t)},
    \label{eq:soliton}
\end{align}
where $\nu$ is a parameter that gives the peak-height of the solution. The numerical evolution of this solution will be discussed in Sec. \ref{sec:ex2} (Example \ref{lst:ex2}).

\subsubsection{Breathers \label{sec:breathers}}
Breathers are periodic solutions along either the $t$-direction, known as Akhmediev Breathers (ABs) \cite{Akhmediev1987a}, or the $x$-direction, known as Kuznetsov-Ma Breathers (KMBs) or Kuznetsov-Ma Solitons \cite{Kuznetsov1977, Ma1979}. This solution can be written as

\begin{align}
   \psi(x,t) = \left[ \frac{(1-4a) \cosh(\delta x) + \sqrt{2a} \cos(\Omega t) + i \delta \sin(\delta x)}{\sqrt{2a} \cos(\Omega t) - \cosh(\delta x)}\right] e^{ix},
\end{align}
where $a$ > 0 is a parameter, $\Omega = 2 \sqrt{1 - 2a}$ is the transverse period for $a < 0.5$ and $\delta = \sqrt{2 a} \Omega$. The values of $0 < a < 0.5$ give Akhmediev breathers, while $a > 0.5$ results in Kuznetsov-Ma breathers. $a \rightarrow 0.5$ gives the spatiotemporally localized Peregrine soliton \cite{Peregrine1983}. The period of the AB is given by $T = 2\pi/\Omega = \pi/\sqrt{1 - 2 a}$.

This solution will be discussed in more detail in Sec. \ref{sec:ex1}, where it is numerically generated from a cosine wave initial condition (Example \ref{lst:ex1}). Higher-order breathers are discussed in Sec. \ref{sec:ex5} (Example \ref{lst:ex5}).

\subsection{Integrals of Motion \label{sec:IoM}}
Being a completely integrable system with infinitely many degrees of freedom, the NLSE has an infinite number of conserved quantities \cite{Faddeev2007, Akhmediev1997}. Here we only list the few lowest-order integrals of motion and refer readers to Miura's original work on the Korteweg-De Vries equation \cite{Miura1968} and Ref. \cite{Nicola1993, Faddeev2007} for an iterative scheme to generate these first integrals.

\begin{align}
\begin{aligned}
N &= \int_{-\infty}^{\infty} |\psi|^2 \diff t, \\
P &= i\int_{-\infty}^{\infty} (\psi_t \psi^* - \psi_t^* \psi) \diff t, \\
H &= \frac{1}{2} \int_{-\infty}^{\infty} (|\psi_t|^2 - |\psi|^4) \diff t = K + V. \\
\end{aligned}
\end{align}

They are the norm, the center of mass momentum, and the Hamiltonian (total energy, which can be split into kinetic and potential terms). We use $K$ to denote the kinetic energy to avoid confusion with the period $T$. For periodic solutions such as Akhmediev breathers, these quantities can be written as
\begin{align}
\begin{aligned}
N &= \frac{1}{T}\int_{0}^{T} |\psi|^2 \diff t, \\
P &= \frac{i}{T}\int_{0}^{T} (\psi_t \psi^* - \psi_t^* \psi) \diff t, \\
H &= \frac{1}{2T} \int_{0}^{T} (|\psi_t|^2 - |\psi|^4) \diff t = K + V. \\
\label{eq:integrals_periodic}
\end{aligned}
\end{align}
These integrals of motion are important because they allow us to check the accuracy of our simulations. We will discuss them in more detail in Sec. \ref{sec:IoM_implementation}, \ref{sec:benchmark} and \ref{sec:ex1} (Example \ref{lst:ex1}).

\section{Solving the NLSE Numerically \label{sec:num_soln}}

\subsection{Splitting Methods \label{sec:splitting}}
Equation \eqref{eq:nlse} can be written as
\begin{align}
    -i \frac{\partial \psi}{\partial x} = (\hat{D} + \hat{N}) \psi.
\end{align}
Here, $\hat{D} = \frac{1}{2} \frac{\partial^2}{\partial t^2}$ is the dispersion operator and $\hat{N} = |\psi|^2$ is the nonlinear operator. To evolve $\psi$ one step $\Delta x$ forward in $x$, we write this formally as
\begin{align}
    \psi(x + \Delta x, t) = e^{\epsilon (\hat{D} + \hat{N})} \psi(x, t) \equiv \mathcal{T}(\epsilon) \psi(x,t), 
\end{align}
where $\epsilon = i \Delta x$. The operator $\mathcal{T}(\epsilon)$ is the evolution operator by an $x$-step $\epsilon$. The Baker-Campbell-Hausdorff (BCH) formula tells us that
\begin{align}
    e^X e^Y = \exp\left(X + Y + \frac{1}{2}[X, Y] + \frac{1}{12} [X, [X,Y]] + \frac{1}{12} [Y, [Y, X] + \ldots \right).
\end{align}
Thus, since $[\hat{D}, \hat{N}] \neq 0$
\begin{align}
    \psi(x + \Delta x, t) \neq e^{\epsilon\hat{D}} e^{\epsilon \hat{N}} \psi(x, t).
\end{align}
However, one can approximate any such operator $e^{\epsilon (\hat{D} + \hat{N})}$ to whichever order in $\epsilon$ one wishes as follows
\begin{align}
e^{\epsilon (\hat{D} + \hat{N})} \approx \prod_{i = 1}^{n} e^{c_i \hat{D} \epsilon} e^{d_i \hat{N} \epsilon},
\end{align}
where $c_i, d_i \in \mathbb{R}$, and for some positive integer $n$. These integrators are symplectic, i.e. they are canonical transformations that preserve the symplectic structure of Hamilton's equations. For a review, see Ref. \cite{Donnelly2005,Chin2020}. One can form two such first order operators by taking $n=1$ and $c_1 = d_1 = 1$
\begin{align}
\begin{aligned}
    \mathcal{T}_{1A}(\epsilon) \equiv e^{\epsilon \hat{N}} e^{\epsilon\hat{D}} , \\
    \mathcal{T}_{1B}(\epsilon) \equiv e^{\epsilon\hat{D}} e^{\epsilon \hat{N}}.
\end{aligned}
\end{align}
However, note that these integrators are not ``time''-reversible (or more accurately with our notation, $x$-reversible). For ``time''-reversible second-order integrators, we can combine $\mathcal{T}_{1A}$ and $\mathcal{T}_{1B}$
\begin{align}
\begin{aligned}
    \mathcal{T}_{2A}(\epsilon) \equiv \mathcal{T}_{1A}(\epsilon/2) \mathcal{T}_{1B}(\epsilon/2) =  e^{\frac{\epsilon}{2} \hat{N}} e^{\epsilon\hat{D}} e^{\frac{\epsilon}{2} \hat{N}}, \\
    \mathcal{T}_{2B}(\epsilon) \equiv \mathcal{T}_{1B}(\epsilon/2) \mathcal{T}_{1A}(\epsilon/2) = e^{\frac{\epsilon}{2} \hat{D}} e^{\epsilon\hat{N}} e^{\frac{\epsilon}{2} \hat{D}}. \\
\end{aligned}
\end{align}
Further, one can show that higher even-order integrators can be formed by combining ones of the preceding order
\begin{align}
    \mathcal{T}_{M,A}(\epsilon) \equiv \mathcal{T}_{M-2,A}(\gamma_1 \epsilon) \mathcal{T}_{M-2,A}(\gamma_2 \epsilon) \mathcal{T}_{M-2,A}(\gamma_3 \epsilon),
    \label{eq:tj}
\end{align}
and similarly for the $B$ variant. Here, $M = 4, 6, 8, \ldots$, and $\gamma_{\{1,2,3\}}$ are some real numbers, for which there are several possible options. One possibility is enforcing ``time''-reversibility ($\gamma_1 = \gamma_3$), and we get \cite{Yoshida1990, Forest1990}
\begin{align}
   \gamma_1 = \gamma_3 =  \frac{1}{2 - 2^{1/(M-1)}}, \quad \gamma_2 =   -\frac{2^{1/(M-1)}}{2 - 2^{1/(M-1)}}.
\end{align}
These are sometimes called the Triple Jump integrators since they involve three applications of the preceding order integrator to get to the desired $x$-step. Another possibility is reducing the size of the steps even further, but this requires a composition of the form
\begin{align}
    \mathcal{T}_{M,A}(\epsilon) \equiv \mathcal{T}_{M-2,A}(\gamma_1 \epsilon) \mathcal{T}_{M-2,A}(\gamma_2 \epsilon)\mathcal{T}_{M-2,A}(\gamma_3 \epsilon)\mathcal{T}_{M-2,A}(\gamma_4 \epsilon) \mathcal{T}_{M-2,A}(\gamma_5 \epsilon).
    \label{eq:suzuki}
\end{align}
One possible ``time''-reversible solution is
\begin{align}
   \gamma_1 = \gamma_2 = \gamma_4 = \gamma_5 =  \frac{1}{4 - 4^{1/(M-1)}}, \quad \gamma_3 =  -\frac{4^{1/(M-1)}}{4 - 4^{1/(M-1)}}.
\end{align}
This family is known as Suzuki's Fractal integrators \cite{Suzuki1990}. However, one does not have to form compositions of the form \eqref{eq:tj} or \eqref{eq:suzuki}, but most generally, they can be written as
\begin{align}
    \mathcal{T}_{M,A}(\epsilon) = \prod_{i}^{N} \mathcal{T}_{2,A}(\gamma_i \epsilon).
\end{align}
Using a numerical approach, many authors managed to solve for the necessary $\gamma_i$ and find such ``optimal'' algorithms to very high-order (i.e., they require far fewer compositions of $\mathcal{T}_2$ than the previous two schemes). For example, one such eighth order algorithm requires only $15$ \cite{Suzuki1993, Suzuki1994, McLachlan1995} applications of $\mathcal{T}_2$, in contrast to $27$ applications of $\mathcal{T}_2$ in the Triple Jump scheme and $125$ in Suzuki's Fractal scheme. These optimal algorithms are discussed in \cite{Yoshida1990, McLachlan1995, Kahan1997, Suzuki1994, Suzuki1993}. 

Another approach by Chin \cite{Chin2010} is through a multi-product decomposition, yielding non-symplectic Runge-Kutta-Nystr\"om (RKN) integrators that require far fewer applications of $\mathcal{T}_2$. 

We have barely scratched the surface of this topic; we direct interested readers to the original papers cited above, the text by Hairer \emph{et al.} \cite{Hairer2006} and Chin's pedagogical review \cite{Chin2020}.

\subsection{Numerical Implementation \label{sec:sim_num}}
This section examines the numerical implementation of the topics discussed above in \texttt{NonlinearSchrodinger.jl}.
\subsubsection{Implementing the Integrators \label{sec:sim_algo_num}}

We will take the algorithm $\mathcal{T}_{1A}$ as an example in this section. To calculate the first part of this integrator, we need to compute
\begin{align}
    \psi_I \equiv e^{\epsilon\hat{N}} \psi(x, t) = e^{i \Delta x |\psi(x,t)|^2} \psi(x,t),
\end{align}
where $\psi_I$ denotes an intermediate step. This is straightforward to compute in real space. For the second half of this integrator, we need to act on $\psi_I$ with the exponential of the dispersion operator $\hat{D}$
\begin{align}
    \psi(x + \Delta x, t) = e^{\epsilon\hat{D}} \psi_I = e^{i \Delta x \frac{1}{2} \partial_t^2} \psi_I,
\end{align}
This operation is much easier to perform in momentum space
\begin{align}
    \psi(x + \Delta x, t) = \mathcal{F}^{-1}\left\{ e^{- i \frac{\Delta x }{2} \omega^2} \mathcal{F} \left\{ \psi_I\right\}, \right\},
\end{align}
where $\mathcal{F}$ and $\mathcal{F}^{-1}$ denote the Fourier and inverse Fourier transforms, respectively. Thus, in total, we get
\begin{align}
    \psi(x + \Delta x, t) = \mathcal{F}^{-1}\left\{ e^{- i \frac{\Delta x }{2} \omega^2} \mathcal{F} \left\{ e^{i \Delta x |\psi(x,t)|^2} \psi(x,t)\right\} \right\}.
\end{align}

Higher-order algorithms can be implemented similarly. The Fourier transforms are performed using the FFT algorithm as implemented in the FFTW library \cite{Frigo2005}. For the exact computational implementation, see the file \texttt{CubicSolvers.jl} in the package.

\begin{figure}
  \sidesubfloat[]{\includegraphics[width=0.9\linewidth]{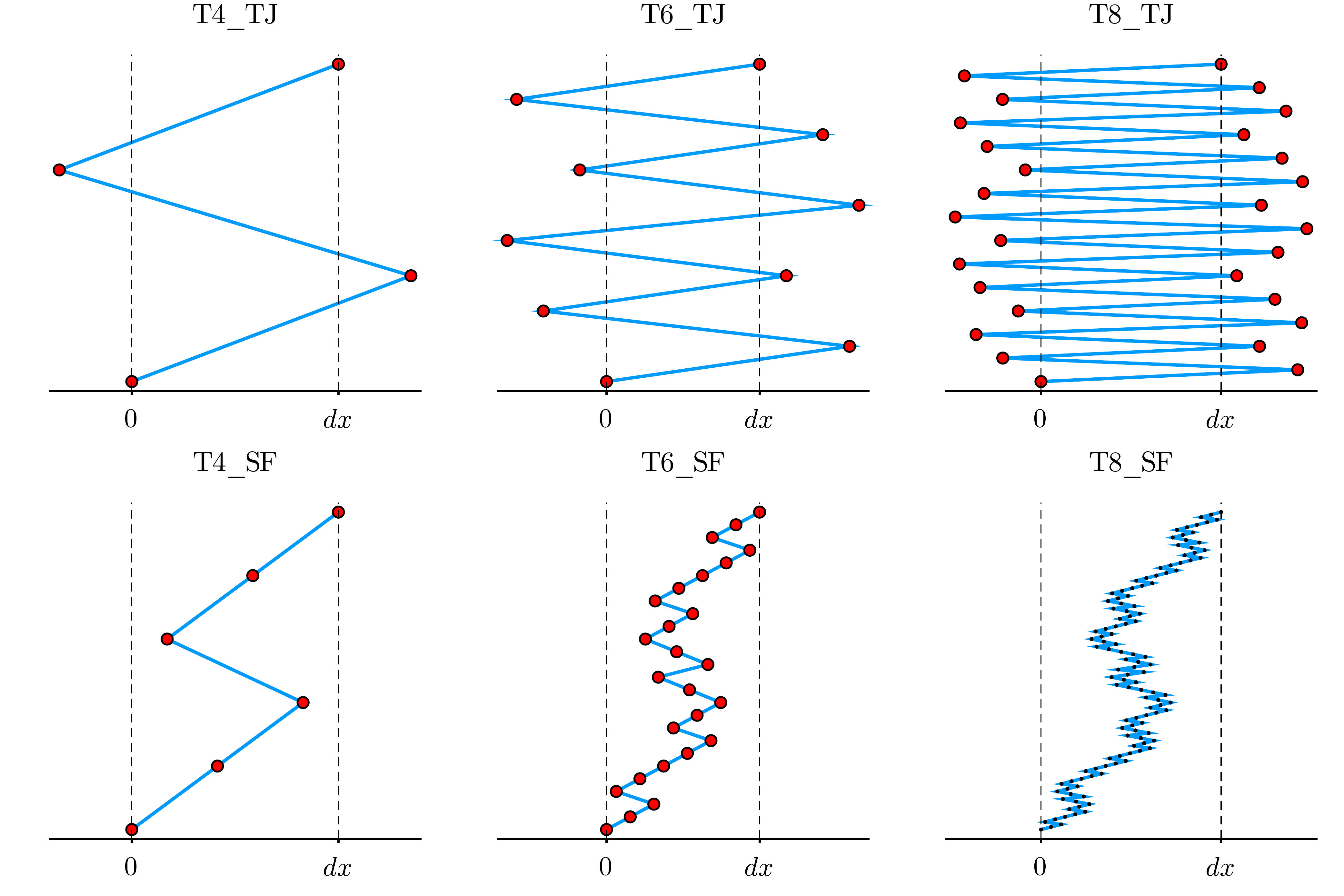}}\\
  \sidesubfloat[]{\includegraphics[width=0.9\linewidth]{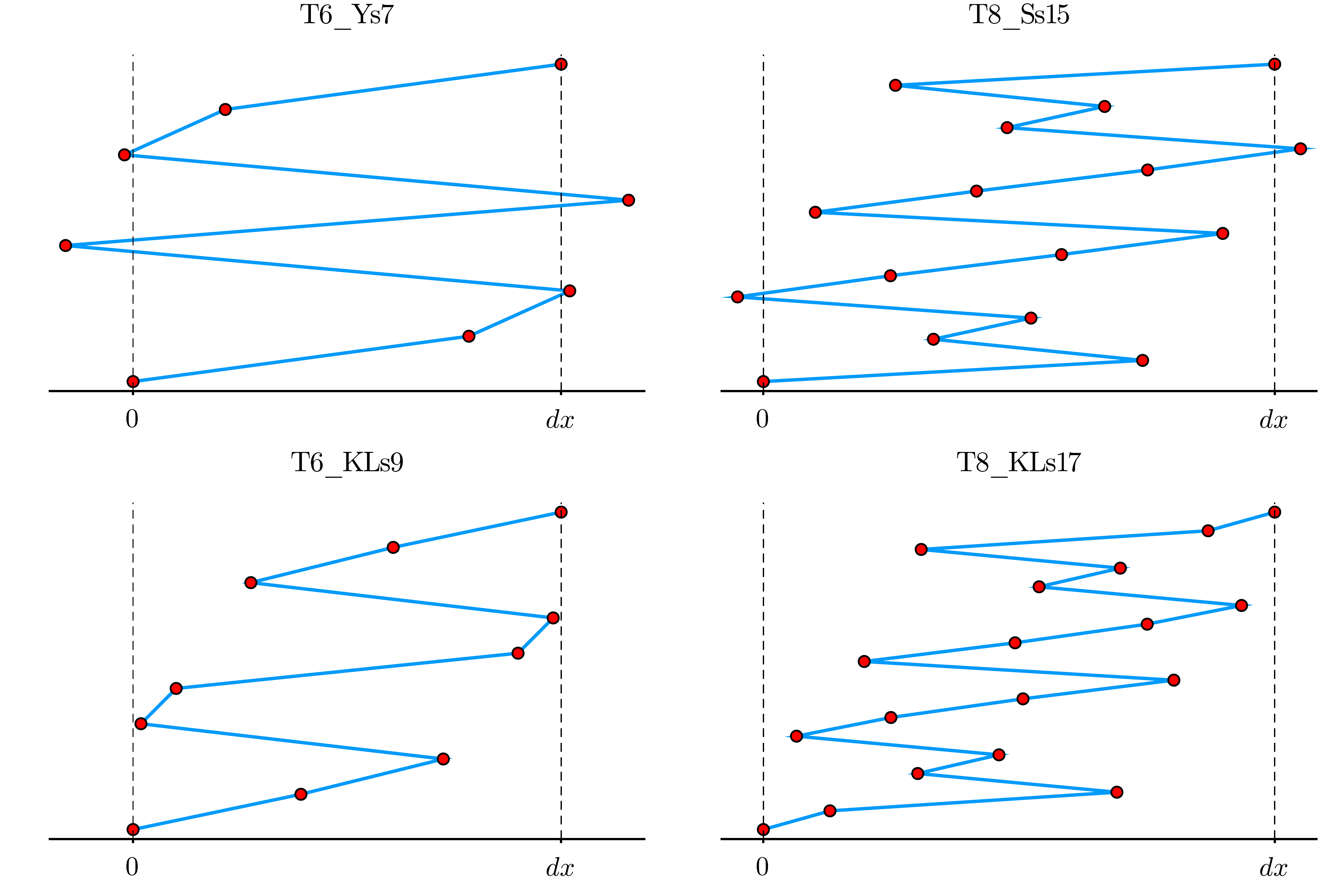}}
  \caption{Visualization of the $x$-stepping of some of the algorithms listed in Table \ref{tab:algorithms}. Since the $A$ and $B$ variants have identical stepping, that label is omitted. \label{fig:algo}}
\end{figure}

Table \ref{tab:algorithms} provides a listing of the algorithms implemented in this package. A visualization of the time-stepping of a selection of these algorithms is shown in Fig. \ref{fig:algo}. 

\begin{table}
\caption{Algorithms available for solving the cubic NLSE in \texttt{NonlinearSchrodinger.jl}. Class $A$ denotes integrators where we evaluate the nonlinear step first, then the dispersion step, and vice versa for class $B$. The parameter $s$, if noted, indicates the number of compositions of the integrator $\mathcal{T}_2$. The number of Fourier transforms (\# FTs) required by each algorithm is multiplied by 2 to account for the inverse Fourier transform.}
\label{tab:algorithms}
\begin{center}
 \begin{tabular}{c c c c c l l} 
 \toprule
 & Function & Order & Type & Description & \# FTs & Ref \\ [0.5ex] 
 \hline
\rownumber & \texttt{T1A!}     & First  & Symplectic & Symplectic Euler $A$      &$2 \times 1$  & \cite{Hairer2006} \\
\rownumber & \texttt{T1B!}     & First  & Symplectic & Symplectic Euler $B$      &$2 \times 1$  & \cite{Hairer2006} \\
\rownumber & \texttt{T2A!}     & Second & Symplectic & Velocity Verlet         &$2 \times 1$  & \cite{Gould2007} \\
\rownumber & \texttt{T2B!}     & Second & Symplectic & Position Verlet         &$2 \times 1$  & \cite{Gould2007} \\
\rownumber & \texttt{T4A\_TJ!} & Fourth & Symplectic & Triple Jump $A$           &$2 \times 3$  & \cite{Forest1990, Yoshida1990}\\
\rownumber & \texttt{T4B\_TJ!} & Fourth & Symplectic & Triple Jump $B$           &$2 \times 3$  &  \cite{Forest1990, Yoshida1990} \\
\rownumber & \texttt{T4A\_SF!} & Fourth & Symplectic & Suzuki's Fractal $A$      &$2 \times 5$  & \cite{Suzuki1990} \\
\rownumber & \texttt{T4B\_SF!} & Fourth & Symplectic & Suzuki's Fractal $B$      &$2 \times 5$  & \cite{Suzuki1990} \\
\rownumber & \texttt{T4A\_CMP!}& Fourth & RKN  & Chin's Multi-Product $A$        &$2 \times 3$  & \cite{Chin2010} \\
\rownumber & \texttt{T4B\_CMP!}& Fourth & RKN  & Chin's Multi-Product $B$        &$2 \times 3$  & \cite{Chin2010} \\
\rownumber & \texttt{T6A\_TJ!} & Sixth & Symplectic & Triple Jump $A$            &$2 \times 9$  & \cite{Yoshida1990} \\
\rownumber & \texttt{T6B\_TJ!} & Sixth & Symplectic & Triple Jump $B$            &$2 \times 9$  & \cite{Yoshida1990} \\
\rownumber & \texttt{T6A\_SF!} & Sixth & Symplectic & Suzuki's Fractal $A$       &$2 \times 25$  & \cite{Suzuki1990} \\
\rownumber & \texttt{T6B\_SF!} & Sixth & Symplectic & Suzuki's Fractal $B$       &$2 \times 25$  & \cite{Suzuki1990} \\
\rownumber & \texttt{T6A\_CMP!}& Sixth & RKN  & Chin's Multi-Product $A$         &$2 \times 6$  & \cite{Chin2010} \\
\rownumber & \texttt{T6B\_CMP!}& Sixth & RKN  & Chin's Multi-Product $B$         &$2 \times 6$  & \cite{Chin2010} \\
\rownumber & \texttt{T8A\_TJ!} & Eighth & Symplectic & Triple Jump $A$           &$2 \times 27$  & \cite{Yoshida1990} \\
\rownumber & \texttt{T8B\_TJ!} & Eighth & Symplectic & Triple Jump $B$           &$2 \times 27$  & \cite{Yoshida1990} \\
\rownumber & \texttt{T8A\_SF!} & Eighth & Symplectic & Suzuki's Fractal $A$      &$2 \times 125$  & \cite{Suzuki1990} \\
\rownumber & \texttt{T8B\_SF!} & Eighth & Symplectic & Suzuki's Fractal $B$      &$2 \times 125$  & \cite{Suzuki1990} \\
\rownumber & \texttt{T8A\_CMP!}& Eighth & RKN  & Chin's Multi-Product $A$        &$2 \times 10$  & \cite{Chin2010} \\
\rownumber & \texttt{T8B\_CMP!}& Eighth & RKN  & Chin's Multi-Product $B$        &$2 \times 10$  & \cite{Chin2010} \\
\rownumber & \texttt{T6A\_Ys7!}& Sixth & Symplectic & Yoshida's $s7$ $A$         &$2 \times 7$  & \cite{Yoshida1990} \\
\rownumber & \texttt{T6B\_Ys7!}& Sixth & Symplectic & Yoshida's $s7$ $B$         &$2 \times 7$  & \cite{Yoshida1990} \\
\rownumber & \texttt{T6A\_KLs9!}& Sixth & Symplectic & Kahan \& Li's $s9$ $A$    &$2 \times 9$  & \cite{Kahan1997} \\
\rownumber & \texttt{T6B\_KLs9!}& Sixth & Symplectic & Kahan \& Li's $s9$ $B$    &$2 \times 9$  & \cite{Kahan1997} \\
\rownumber & \texttt{T6A\_Ss14!}& Sixth & Symplectic & Suzuki's $s14$ $A$        &$2 \times 14$  & \cite{Suzuki1994} \\
\rownumber & \texttt{T6B\_Ss14!}& Sixth & Symplectic & Suzuki's $s14$ $B$        &$2 \times 14$  & \cite{Suzuki1994} \\
\rownumber & \texttt{T8A\_Ss15!}& Eighth & Symplectic & Suzuki's $s15$ $A$       &$2 \times 15$  & \cite{Suzuki1993, Suzuki1994, McLachlan1995}\\
\rownumber & \texttt{T8B\_Ss15!}& Eighth & Symplectic & Suzuki's $s15$ $B$       &$2 \times 15$  & \cite{Suzuki1993, Suzuki1994, McLachlan1995} \\
\rownumber & \texttt{T8A\_KLs17!}& Eighth & Symplectic & Kahan \& Li's $s17$ $A$ &$2 \times 17$  & \cite{Kahan1997}\\
\rownumber & \texttt{T8B\_KLs17!}& Eighth & Symplectic & Kahan \& Li's $s17$ $B$ &$2 \times 17$  & \cite{Kahan1997} \\
 \bottomrule
\end{tabular}
\end{center}
\end{table}

\subsubsection{Computing the Integrals of Motion and Errors \label{sec:IoM_implementation}}
\texttt{NonlinearSchrodinger.jl} allows the computation of the first 3 integrals of motion of the NLSE, shown in Eq. \eqref{eq:integrals_periodic}. The norm and potential energy are straightforward to calculate numerically in real space. However, we perform $t$ derivatives in $\omega$-space when computing the momentum and kinetic energy integrals

\begin{align*}
    K &=  \frac{1}{2T} \int_{0}^{T} |\psi_t|^2 \diff t = \frac{1}{T} \int_0^T \psi^*(x,t) \left(\frac{-1}{2} \partial_x^2\right) \psi(x,t) \diff t \\
      &= \frac{-1}{2T} \sum_{\omega}\sum_{\omega'}\int_{0}^{T} \tilde{\psi}_\omega^*(x) e^{+i \omega t} (- i \omega'^2) e^{-i \omega' t}\tilde{\psi}_\omega(x) \diff t \\
      &= \frac{1}{2} \sum_\omega \omega^2 |\tilde{\psi}_\omega(x)|^2, \\
    P &= \frac{i}{T}\int_{0}^{T} (\psi^* \psi_t  - \psi_t^* \psi) \diff t = \frac{-2}{T}\int_0^T\mathfrak{Im}\left\{\psi^* \psi_t\right\} \\
      &= \frac{-2}{T} \int_{0}^{T} \mathfrak{Im} \left\{\sum_{\omega} \sum_{\omega'} e^{i (\omega - \omega')t} (- i \omega') \tilde{\psi}^*_\omega(x) \tilde{\psi}_{\omega'}(x)\right\} \\
      &= 2 \sum_\omega \mathfrak{Im}\left\{ i \omega |\tilde{\psi}_\omega(x)|^2 \right\} = 2 \sum_\omega \omega |\tilde{\psi}_\omega(x)|^2.
\end{align*}
Here, we use $^*$ to denote complex conjugation and $\tilde{\psi}_\omega(x)$ is the Fourier component of $\psi(x,t)$ with frequency $\omega$. In conclusion, the integrals of motion are computed as follows
\begin{align}
    \begin{aligned}
    N(x) &= \frac{1}{T}\int_{0}^{T} |\psi|^2 \diff t, \\
    P(x) &= 2 \sum_\omega \omega |\tilde{\psi}_\omega(x)|^2 , \\
    K(x) &= \frac{1}{2} \sum_\omega \omega^2 |\tilde{\psi}_\omega(x)|^2, \\
    V(x) &= \frac{-1}{2T} \int_{0}^{T} |\psi|^4 \diff t,  \\
    E(x) \equiv H(x) &= K(x) + V(x).\\
    \end{aligned}
\end{align}
Their errors are simply defined as
\begin{align}
   \delta F(x) \equiv F(x) - F(x=0)
   \label{eq:error}
\end{align}
where $F = N, \,P, \, E$. These are computed via the function \texttt{compute\_IoM!}, which will be demonstrated in Sec. \ref{sec:ex1} (Example \ref{lst:ex1}).

\subsection{Benchmarking the Algorithms \label{sec:benchmark}}

To benchmark the algorithms presented in Sec. \ref{sec:sim_num} (Table \ref{tab:algorithms}), we perform multiple simulations with a fixed length and a different value of $N$, where $N$ denotes the number of applications of $\mathcal{T}_2$ (i.e., half the number of Fourier transforms). Since the total length of the simulation is fixed, $dx$ decreases as $N$ increases. We use a cosine wave initial condition \eqref{eq:cosineseries}, which is discussed in more detail in Sec. \ref{sec:ex1} (Example \ref{lst:ex1}). For each simulation, we compute the energy error \eqref{eq:error} to gauge the performance of the different algorithms.

\begin{figure}
  \sidesubfloat[]{\includegraphics[width=0.9\linewidth]{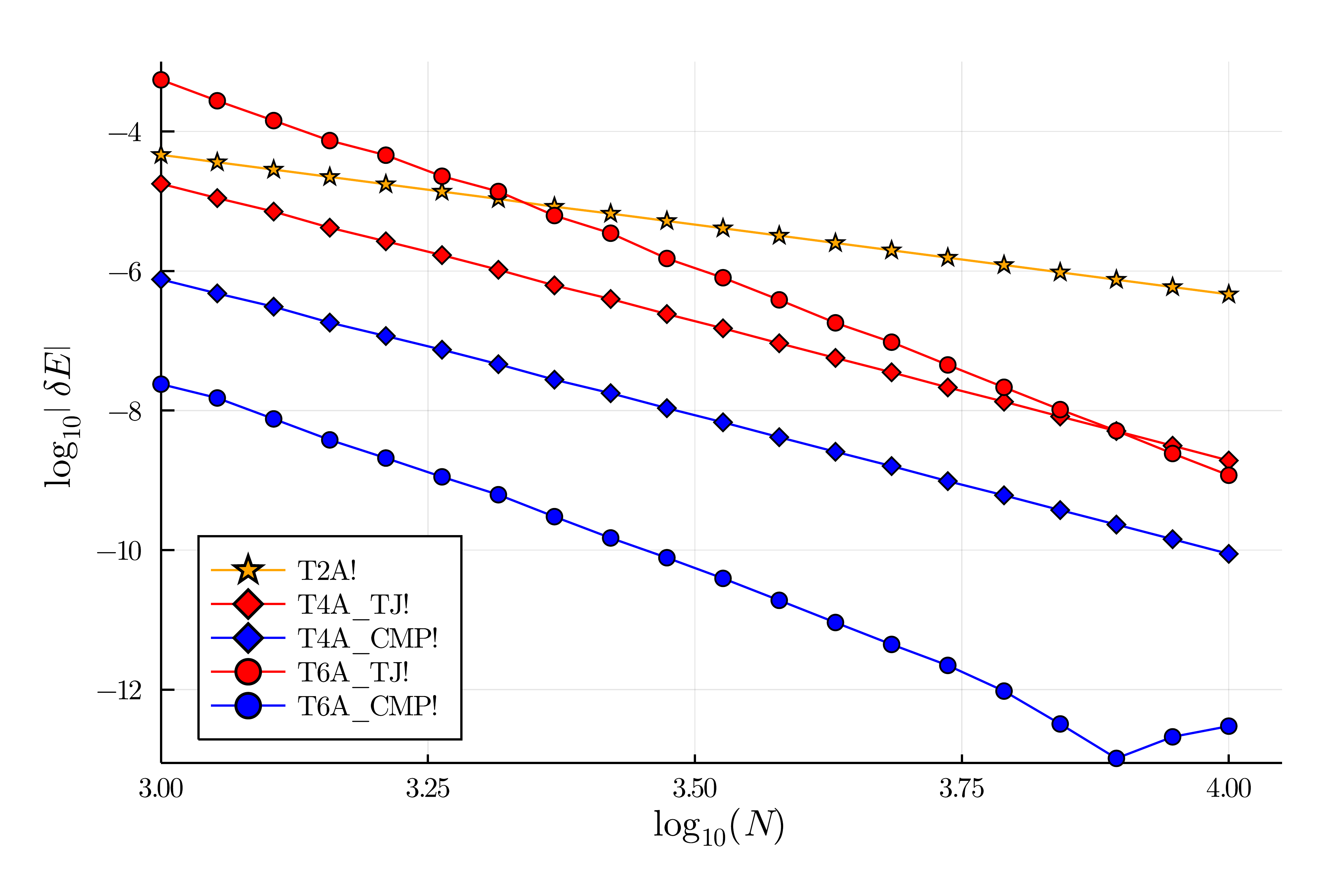} \label{fig:error_246}}\\
  \sidesubfloat[]{\includegraphics[width=0.9\linewidth]{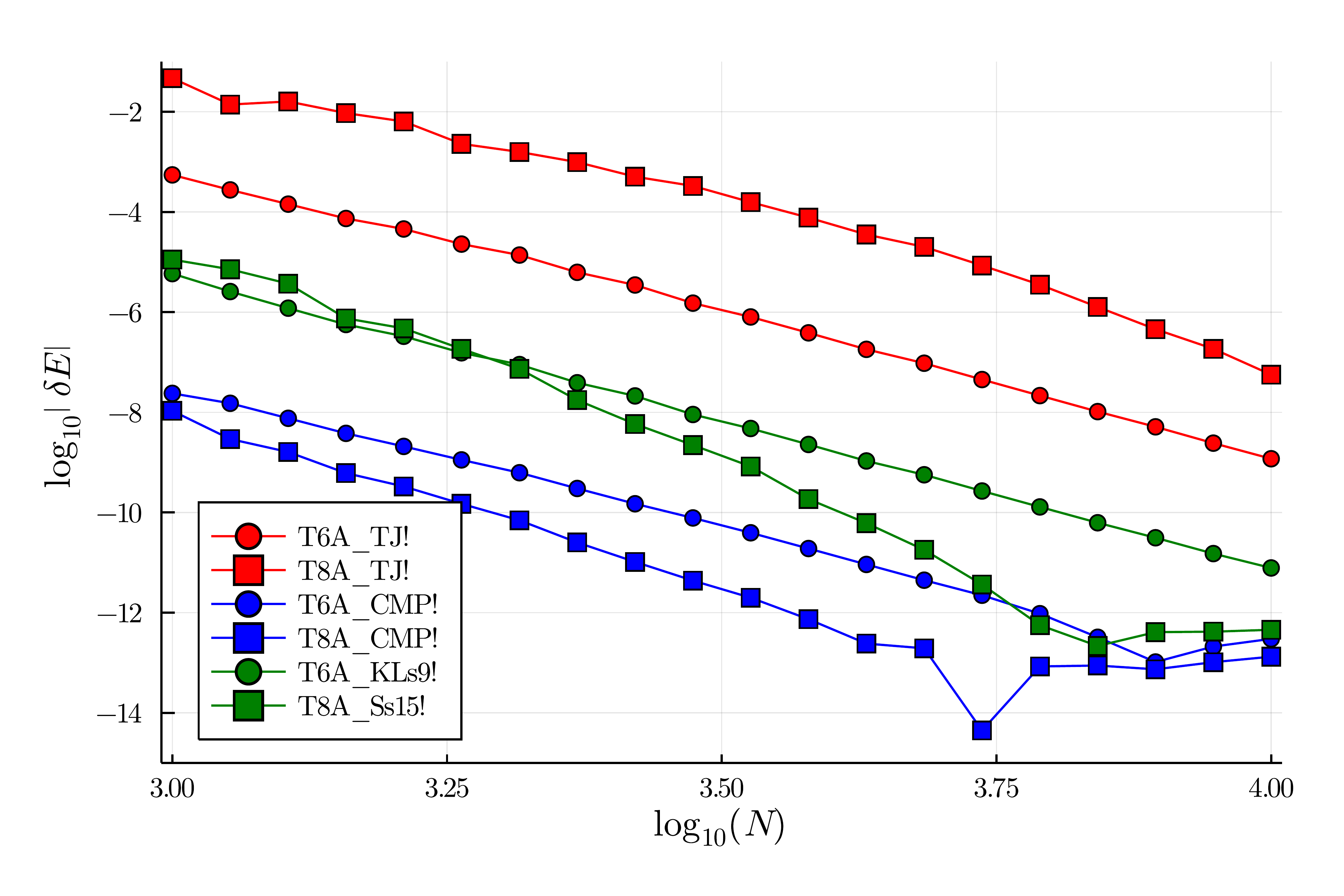} \label{fig:error_68}}
  \caption{The energy error \eqref{eq:error} versus $N$ for a selection of algorithms. Algorithms of orders two, four, six and eight are shown as stars, diamonds, circles and squares, respectively. Each family has a different color, e.g. the Triple Jump family (TJ) is all shown in red. In (a), we show lower-order integrators, up to sixth order, and in (b), we show higher-order integrators of orders six and eight. For a description of the algorithms, refer to Table \ref{tab:algorithms}. This plot specifically shows the $A$ class of algorithms. Similar, albeit not identical, results can be computed for the $B$ class. \label{fig:error}}
\end{figure}

A plot of the energy error \eqref{eq:error} versus $N$ is shown in Fig. \ref{fig:error} for a selection of algorithms. As demonstrated in Fig. \ref{fig:error_246}, all benchmarked algorithms of orders two through six display a linear dependence of $\log_{10}|\delta E|$ on $\log_{10}(N)$, as expected. The higher the order of the integrator, the steeper the slope. 

The second-order symplectic integrator \texttt{T2A!} performs quite poorly, with a comparatively large energy error. Moreover, we can see that the Triple Jump family does not perform well either, with the sixth order integrator (\texttt{T6A\_TJ!}) overtaking the fourth-order one (\texttt{T4A\_TJ!}) only at large $N$ (i.e., small $dx$). Chin's Multi-Product family (CMP) performs the best, with \texttt{T6A\_CMP!} having the smallest energy error in Fig. \ref{fig:error_246}. More generally, it is the most accurate sixth-order integrator in the package.

On the other hand, in Fig. \ref{fig:error_68}, we can see that the eighth order Triple Jump algorithm (\texttt{T8A\_TJ!}) does not perform particularly well, with a slight nonlinear dependence of $\log_{10}|\delta E|$ on $\log_{10}(N)$. This nonlinear dependence is an artifact of double-precision coupled with the 27 evaluations of $\mathcal{T}_2$ necessary per step of \texttt{T8A\_TJ!}. In contrast, the optimal integrator \texttt{T8A\_Ss15!} displays a mostly linear behavior, outperforming the sixth order optimal algorithm \texttt{T6A\_KLs9!} quite quickly, at a relatively large $dx$. Still, Chin's Multi-Product family is the best performer; it demonstrates a significantly smaller energy error than all other integrators. 

One can see that the algorithms are bottlenecked by double-precision at around $\delta E \sim 10^{-14}$ and $N \geq 10^{3.75}$. To properly benchmark these eighth-order integrators without artifacts, one must implement quadruple precision, which is part of our road map as per \ref{sec:conc}.

Overall, we recommend using Chin's Multi-Product integrators (CMP) when symplecticity is not a concern; they are the quickest and most accurate integrators this package has to offer. When one wishes to use symplectic integrators, we recommend Kahan and Li's algorithms, \texttt{T6\_KLs9!} and \texttt{T8\_KLs17!}, as they offer the lowest error and are almost as quick as the nearest symplectic competitors.

\subsection{The Necessity of Higher-Order Integrators \label{sec:convergence}}

Suppose we want to solve \eqref{eq:nlse} numerically using the initial condition

\begin{align}
    \psi(x = 0, t) = e^{i \pi t/10} (1 + 0.002 \cos(\pi t /10)).
    \label{eq:bad_ic}
\end{align}

\begin{figure}
  \sidesubfloat[]{\includegraphics[width=0.43\linewidth]{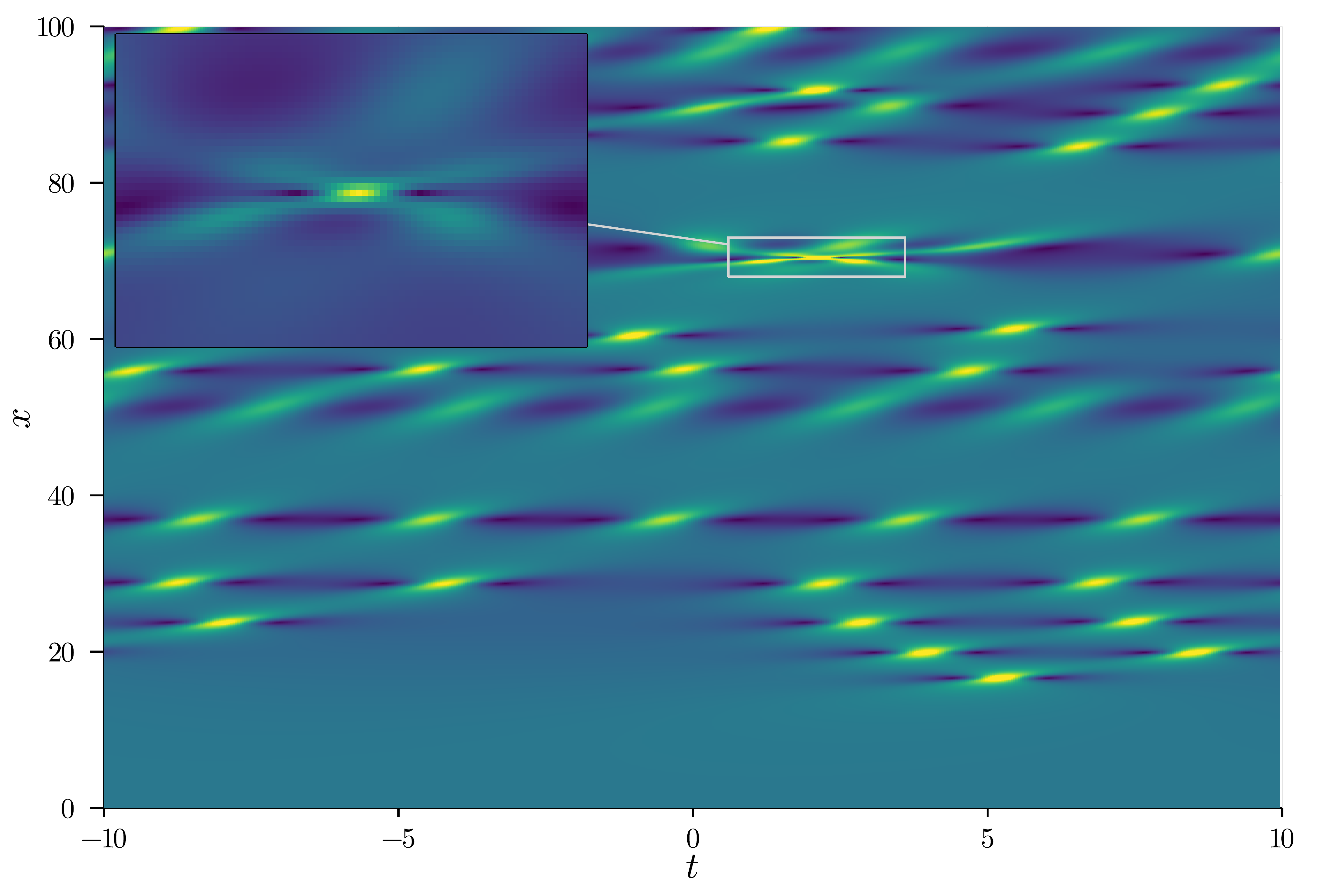}\label{fig:convg2}} \quad
  \sidesubfloat[]{\includegraphics[width=0.43\linewidth]{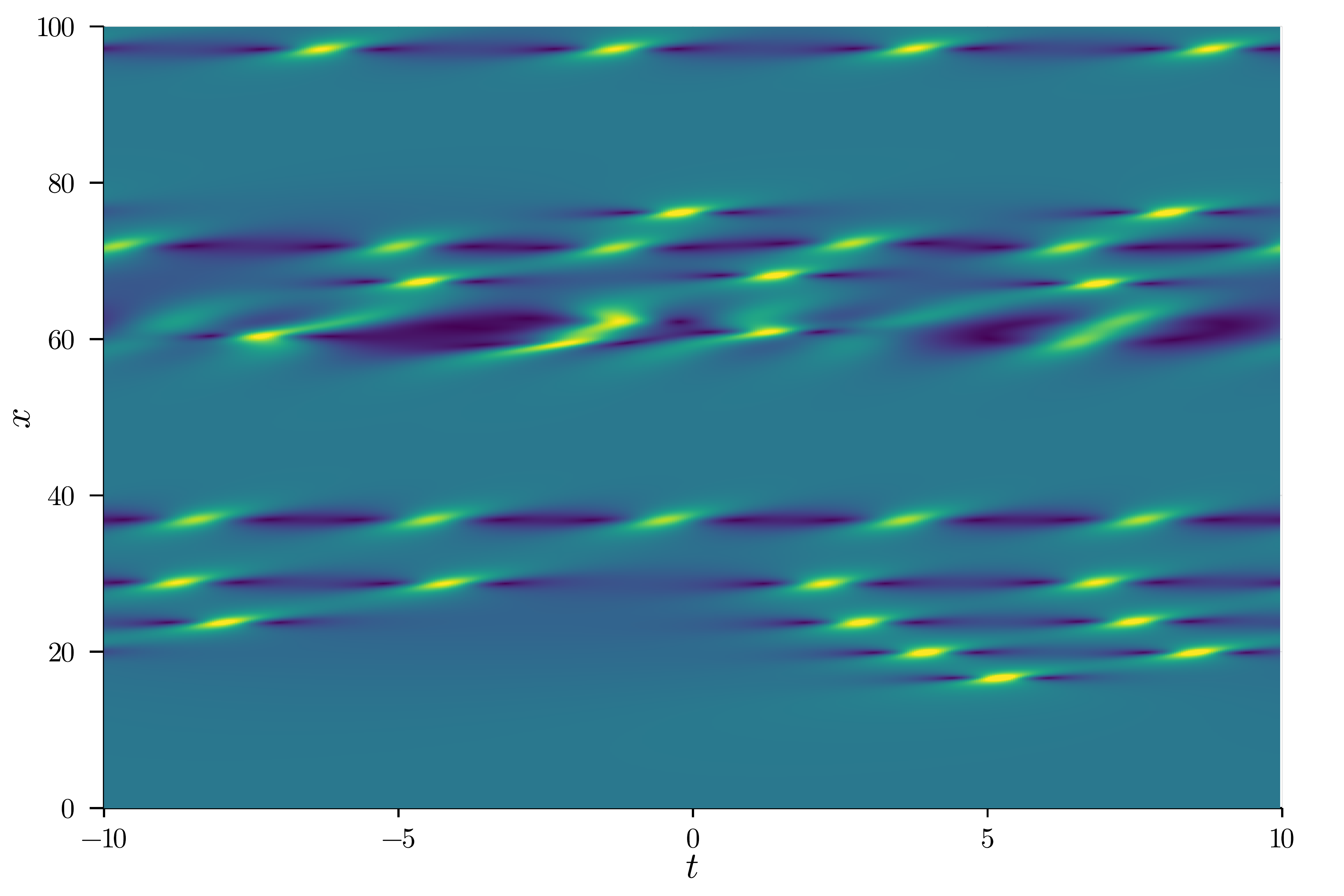}\label{fig:convg4}} \\
  \sidesubfloat[]{\includegraphics[width=0.43\linewidth]{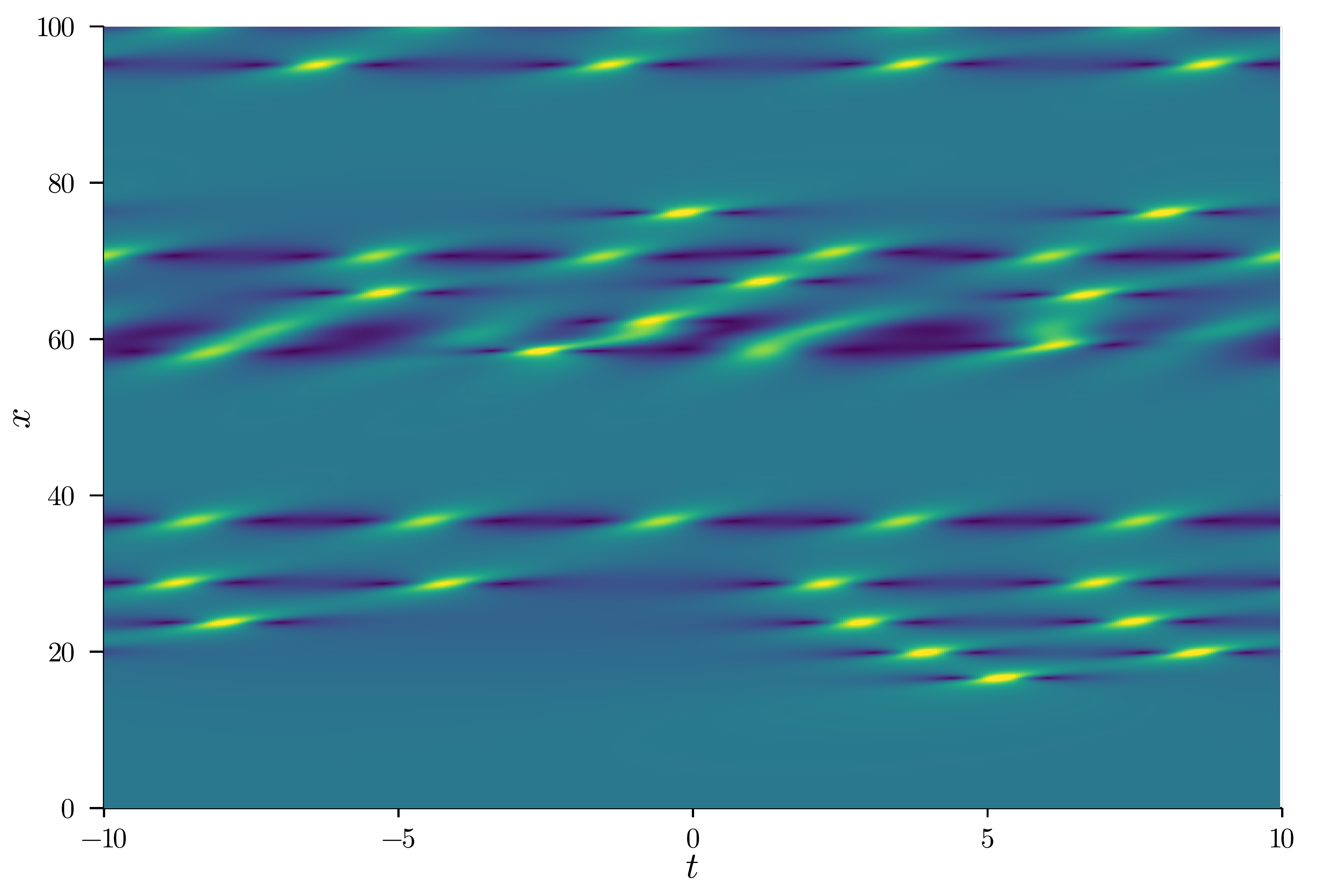}\label{fig:convg6}} \quad
  \sidesubfloat[]{\includegraphics[width=0.43\linewidth]{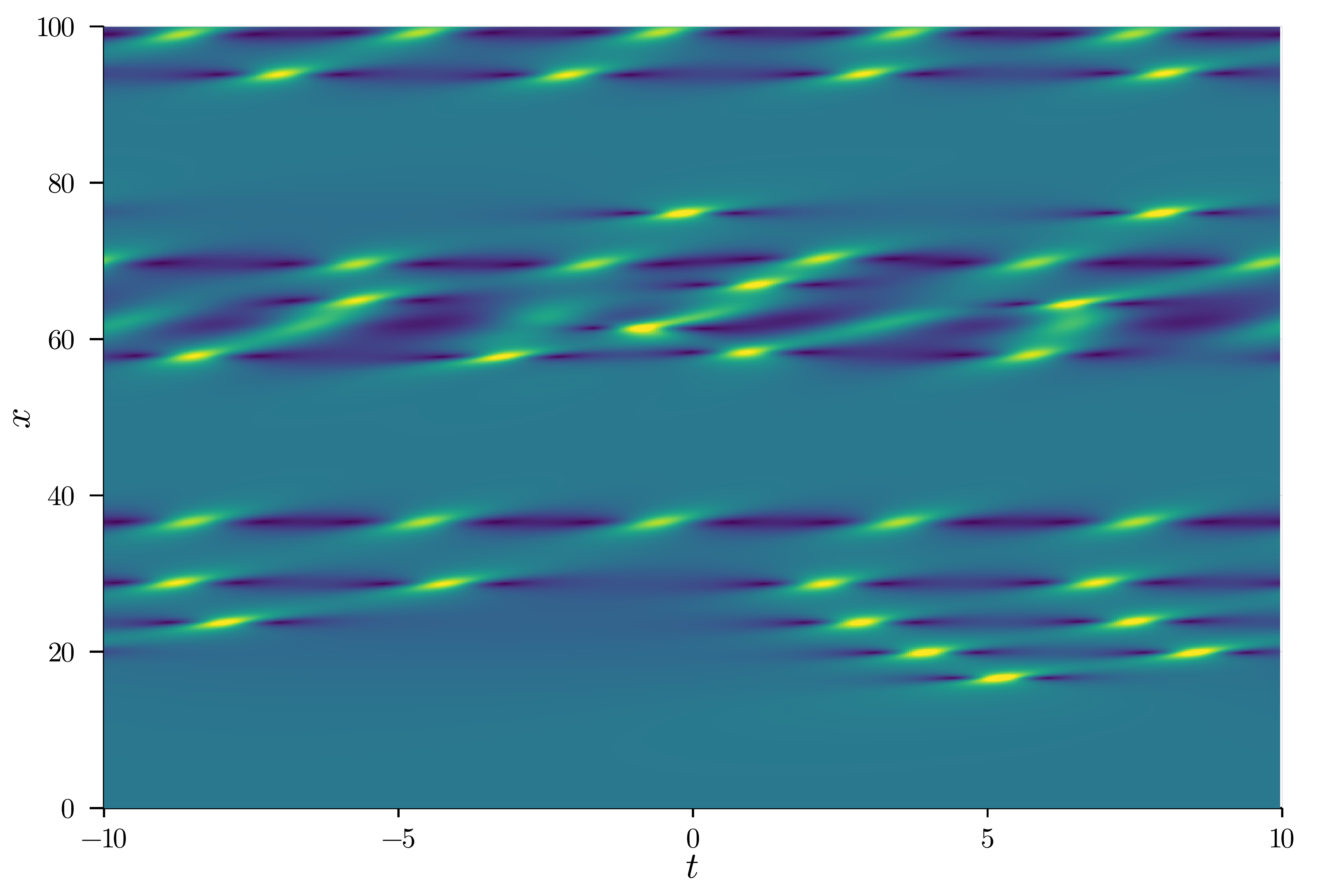}\label{fig:convg8}}
  \caption{Solving \eqref{eq:nlse} numerically using the initial condition \eqref{eq:bad_ic} with $\Delta x = 10^{-4}$ and $N_t = 512$. The algorithms are (a) \texttt{T2A!} (b) \texttt{T4\_TJ!} (c) \texttt{T6\_KLs9!} (d) \texttt{T8\_KLs17!}. Refer to Table \ref{tab:algorithms} for more details. The inset in (a) shows the higher-order peak, which has a height of 5.69. The lower-order peaks have a height of 2.98.\label{fig:conv}}
\end{figure}

The purpose of this section is to demonstrate that higher-order integrators are crucial and that the second-order algorithm is generally insufficient. In Fig. \ref{fig:convg2}, using a second-order symplectic integrator, we see a localized peak much higher than the background, what seems to be a high-amplitude rogue wave. However, upon using higher-order integrators, this peak immediately vanishes. Furthermore, the peak disappears upon halving the $x$-step, using the same second-order integrator (results not shown).

Moreover, it can be seen that the calculations performed using fourth and sixth order integrators (Fig. \ref{fig:convg4} and \ref{fig:convg6}) agree well up to $x \sim 40$. In contrast, sixth and eighth order integrators (Fig. \ref{fig:convg6} and \ref{fig:convg8}) are almost identical, except at $x=95$ where they slightly diverge. Hence, it seems that the calculation is converging using an eighth order integrator at this $x$-step. However, more checking must be done using a smaller step (or a higher-order integrator, preferably with quadruple-precision), which is not shown here.

In conclusion, the first step towards convergence is simply checking the solution after reducing the $x$-step or using a higher-order integrator, even before examining the error in the integrals of motion. In this way, one can avoid making hasty conclusions about extreme events such as rogue waves or higher-order breathers, which are very unlikely to spuriously appear using an arbitrary initial condition without finely tuned parameters (see Ref. \cite{Chin2016}).

\subsection{Using \texttt{NonlinearSchrodinger.jl} for simulations \label{sec:using_nlss_for_sim}}
This section provides three examples for using the package to perform simulations of the cubic NLSE \eqref{eq:nlse}. These examples are for demonstration purposes only and are designed to run quickly on most machines; they are not fully converged calculations. We recommend consulting Appendix \ref{app:install} before reading this section for users unfamiliar with Julia.

\subsubsection{Example 1: Cosine Wave initial condition \label{sec:ex1}}
In this section, we use a cosine wave initial condition to ``seed'' an Akhmediev breather, leading to its recurrence via modulation instability. See Ref. \cite{Chin2015} and the references therein for more details. This initial condition is given by
\begin{align}
    \psi_0 \equiv \psi(x=0, t) = A_0 + 2 \sum_{1}^{n} A_m \cos(m \Omega t),
    \label{eq:cosineseries}
\end{align}
for some real or complex coefficients $A_m$. As before, $\Omega = 2\sqrt{1 - 2 a}$ denotes the frequency of the breather, and its period is $T = 2\pi/\Omega$. We will often use the complex eigenvalue $\lambda = i \nu = i \sqrt{2 a}$ in lieu of the parameter $a$, such that $\lambda \in (0, 1)$. We can always fix $A_0$ to be real, and by the normalization condition
\begin{align}
N = 1 = \frac{1}{T} \int_0^T |\psi_0|^2 \diff t = A_0^2 + 2 \sum_{m=1}^n |A_m|^2 \implies A_0 = \sqrt{1 - 2 \sum_{m=1}^n |A_m|^2}.
\end{align}

To run such a simulation, we must select one of the three necessary parameters, $\lambda$ (or equivalently, $a$), $\Omega$, or $T$, then we compute the remaining two
\begin{lstlisting}[language=Julia]
λ, T, Ω = params(λ = 0.8im)   # or
λ, T, Ω = params(a = 0.8^2/2) # or
λ, T, Ω = params(Ω = 1.2)     # or
λ, T, Ω = params(T = 2*Pi/1.2)
\end{lstlisting}

The user should pick one of the four options; they are equivalent. However, note that everything is done in terms of $\lambda$, not $a$. $a$ is only used as a parameter for breathers, while $\lambda$ is used in the literature for other solutions as well.

The next step is to generate a simulation box. Since we are studying a periodic solution, the box's transverse size should be a multiple of the breather's period. For one period $T$, we have $t \in [-T/2, T/2)$. Any longitudinal range of $x$ can be selected. The number of nodes in the transverse direction $N_t$ should also be specified, as well as the longitudinal grid spacing, $dx$. Note that $N_t$ is the number of grid points used in the computation of the Fourier transform.

\begin{lstlisting}[language=Julia]
xᵣ = 0=>100
box = Box(xᵣ, T, dx=1e-3, Nₜ=256)
\end{lstlisting}

Next, we generate the initial condition \eqref{eq:cosineseries} using a helper function that automates the process.
\begin{lstlisting}[language=Julia]
coeff = [1e-4]
ψ₀, A₀ = ψ₀_periodic(coeff, box, Ω)
\end{lstlisting}
The size of the array \texttt{coeff} = $A_m$ is arbitrary, and the user can specify as many coefficients as they wish; we use only one in this example for simplicity. The final step is to create a simulation structure utilizing the box and initial condition we have produced, as well as $\lambda$ and an algorithm from Table \ref{tab:algorithms}. Then we can solve it and (optionally) compute the integrals of motion.
\begin{lstlisting}[language=Julia]
sim = Sim(λ, box, ψ₀, T4A_TJ!) # Create simulation structure
solve!(sim)                    # Solve the simulation
compute_IoM!(sim)              # Compute the integrals of motion
\end{lstlisting}

At this point, the simulation has concluded, and one can access the solution as follows

\begin{lstlisting}[language=Julia]
sim.ψ   # Array containing the solution ψ(x,t)
sim.ψ̃   # Array containing the spectrum of the solution ψ̃(ω,x)
sim.KE  # Array containing the kinetic energy K(x)
sim.PE  # Array containing the potential energy V(x)
sim.E   # Array containing the energy E(x)
sim.N   # Array containing the norm N(x)
sim.P   # Array containing the momentum P(x)
\end{lstlisting}

If one wishes to visualize the solution, \texttt{NonlinearSchrodinger.jl} provides \texttt{Plots.jl} ``recipes'' that allow normal plotting commands to interpret \texttt{Sim} (and \texttt{Calc}, discussed later) objects. An example is shown below. \footnote{3D surface plots in this manuscript were produced using different plotting software, not these automated recipes. All other plots of the solutions were produced using the recipes.} 
\vspace{5pt}
\begin{lstlisting}[language=Julia]
using Plots

surface(sim)                           # plot |ψ| in surface mode
contour(sim)                           # plot |ψ| in contour mode
heatmap(sim, res_x=1000, res_t=1000)   # plot |ψ| in heatmap mode 
                                       # using 1000 x nodes and 1000 t nodes
                                       # Default resolution is 500x500
heatmap(sim, :ψ̃)                       # plot log|ψ̃| in heatmap mode
plot(sim, :ψ̃)                          # plot log|ψ̃| in lines mode
plot(sim, :IoM)                        # plot integrals of motion and errors 
\end{lstlisting}

These commands function as any \texttt{Plots.jl} commands would, and accept keyword arguments to adjust the plot parameters. All standard plotting commands should work; see the \texttt{Plots.jl} documentation for more details.

\begin{figure}
  \sidesubfloat[]{\includegraphics[width=0.9\linewidth]{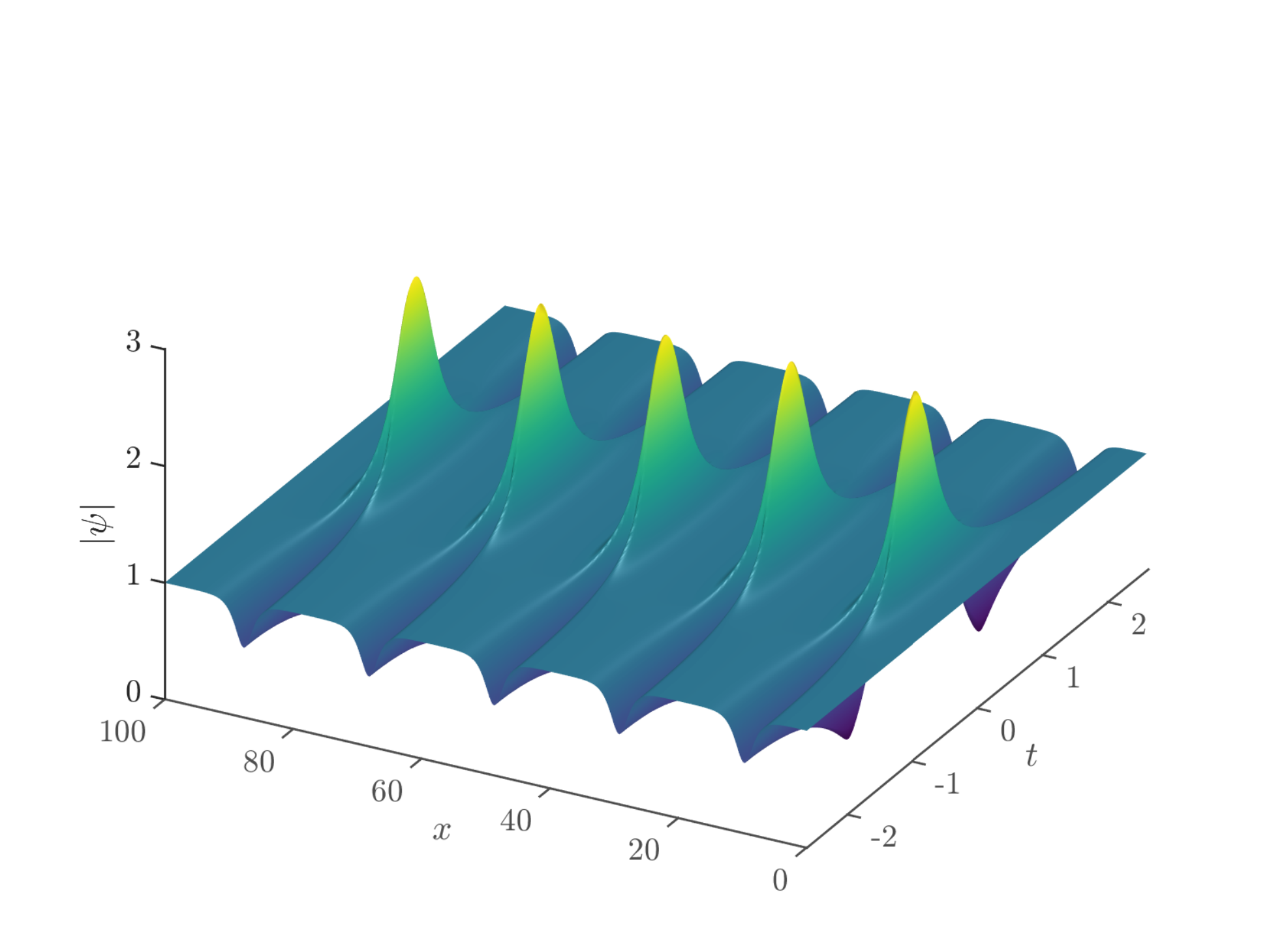}\label{fig:cosine_psi}}\\
  \sidesubfloat[]{\includegraphics[width=0.9\linewidth]{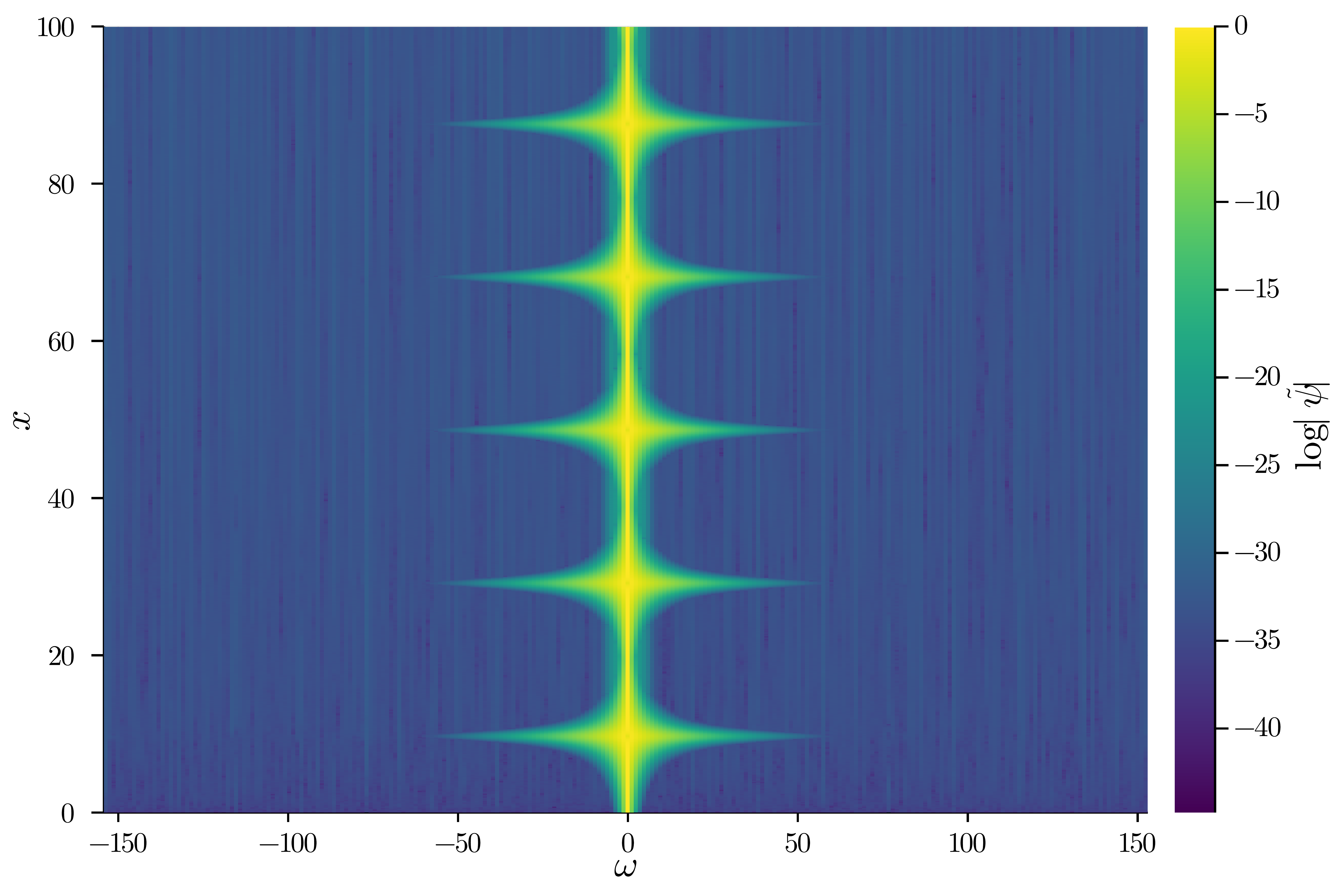}\label{fig:cosine_spectrum}}
  \caption{The results of Example \ref{lst:ex1}, with a cosine wave \eqref{eq:cosineseries} initial condition. (a) The absolute value of the wave envelope, $|\psi(x, t)|$, showing the breather recurrence due to modulation instability. (b) The log of the spectrum of the solution, showing the Fourier modes' growth, the spectral signature of modulation instability. \label{fig:breathers}}
\end{figure}

\begin{figure}
    \centering
    \includegraphics[width=1.0\linewidth]{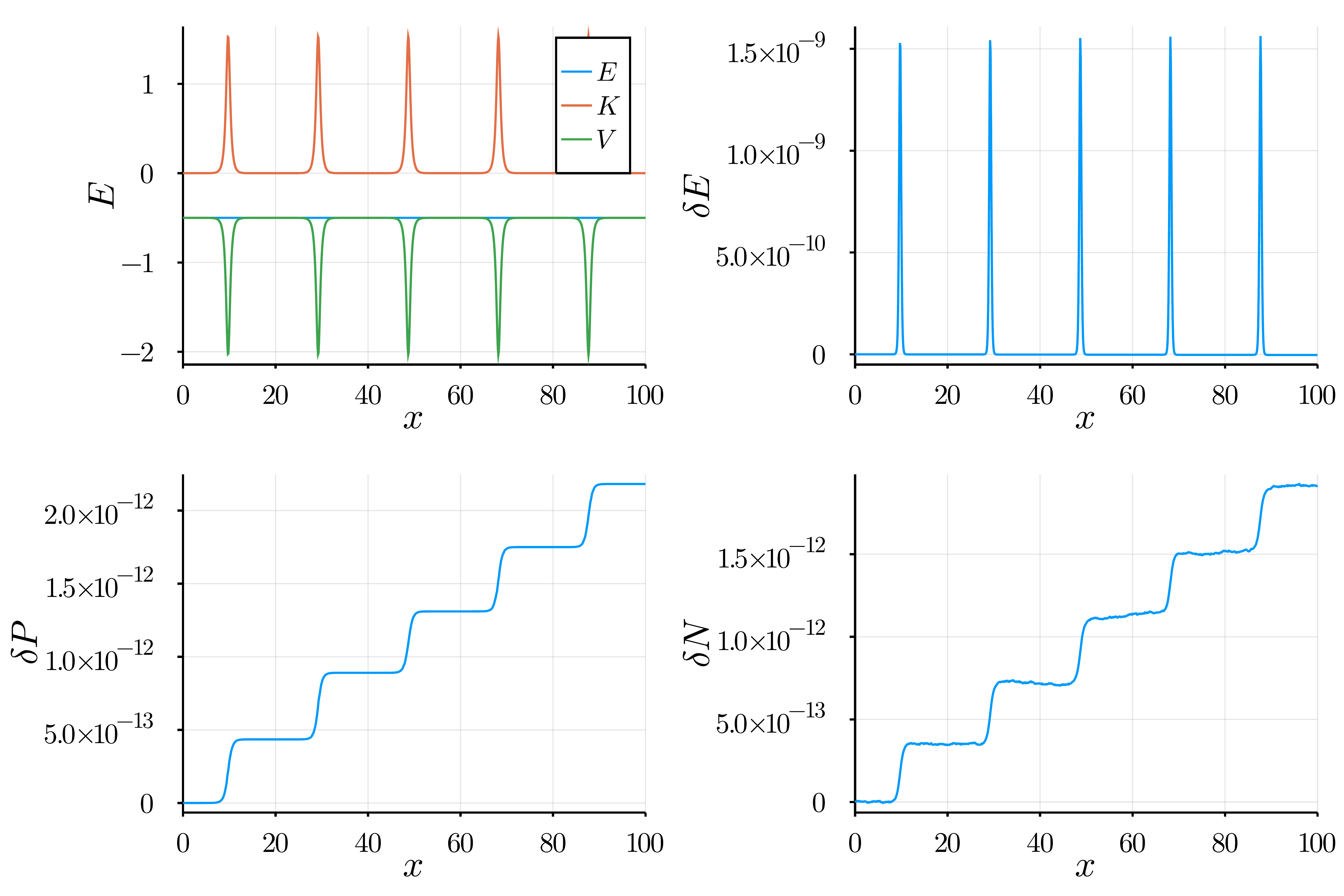}
    \caption{The energy (total energy $E$, kinetic energy $K$ and potential energy $V$), energy error, momentum error and norm error in Example \ref{lst:ex1}, corresponding to a cosine wave \eqref{eq:cosineseries} initial condition. The energy error peaks when the breather forms.}
    \label{fig:cosine_IoM}
\end{figure}

The full example is shown in listing \ref{lst:ex1}, showcasing the simplicity of \texttt{NonlinearSchrodinger.jl}'s interface. Its results are displayed in Fig. \ref{fig:breathers} and \ref{fig:cosine_IoM} \footnote{All examples can also be found in the package's online documentation at \url{https://oashour.github.io/NonlinearSchrodinger.jl/stable/}, or as a Jupyter notebook in the examples folder}.

\begin{lstlisting}[language=Julia, float, floatplacement=H!, caption={Running a simulation with a cosine wave initial condition.}, captionpos=b, label={lst:ex1}]
λ, T, Ω = params(λ = 0.8im)

xᵣ = 0=>100
box = Box(xᵣ, T, dx=1e-3, Nₜ = 256, n_periods = 1)

coeff = [1e-4]
ψ₀, A₀ = ψ₀_periodic(coeff, box, Ω)

sim = Sim(λ, box, ψ₀, T4A_TJ!)

solve!(sim)
compute_IoM!(sim)
\end{lstlisting}

\subsubsection{Example 2: Soliton Initial Condition \label{sec:ex2}}
We are not restricted to the helper function $\psi_0$\texttt{\_periodic} to generate the initial condition. In fact, we can use any initial condition as long as it is an array of type \texttt{Complex\{Float64\}} and has the correct size (\texttt{N}$_t$). For example, we can use the soliton \eqref{eq:soliton} at $x=0$ as an initial condition, as shown in Example \ref{lst:ex2}.

\begin{figure}
    \centering
    \includegraphics[width=0.9\linewidth]{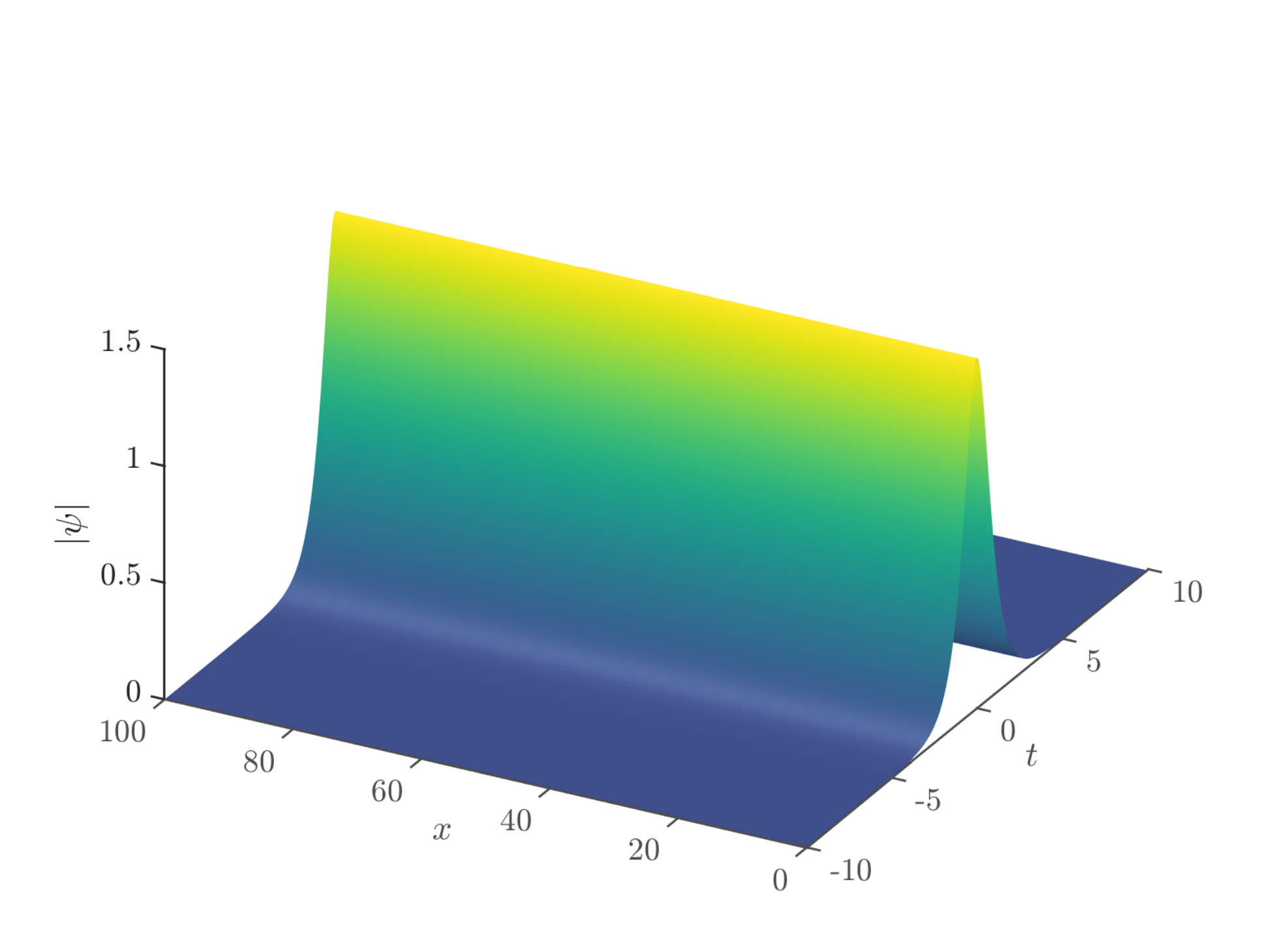}
    \caption{Simulation with a soliton initial condition (Example \ref{lst:ex2}). The soliton maintains its shape while propagating.}
    \label{fig:soliton}
\end{figure}

Due to the soliton's non-periodicity and the inherent periodic boundary conditions of the Fourier split-step algorithms, one must select a large enough transverse box size to ensure the solition decays sufficiently at the boundaries of the box. As shown in Fig. \ref{fig:soliton}, the simulation propagates the soliton without changing its shape, as expected.

\begin{lstlisting}[language=Julia, float, floatplacement=H!, caption={Running a simulation with a soliton initial condition.}, captionpos=b, label={lst:ex2}]
λ = 0.75im

T = 20
xᵣ = 0=>100
box = Box(xᵣ, T, dx=1e-3, Nₜ = 256, n_periods = 1)

ψ₀ .= 2*imag(λ)./cosh.(2*imag(λ).*box.t) .+ 0*im

sim = Sim(λ, box, ψ₀, T4A_TJ!)

solve!(sim)
\end{lstlisting}

\subsubsection{Example 3: Pruning and Nonlinear Talbot Carpets \label{sec:ex3}}

Assume one would like to run a simulation with $M$ periods and $N_t$ Fourier modes. In this case, we expect that the $M^\text{th}, 2 M^\text{th}, \ldots, N_t M^\text{th}$ modes to grow together in lock-step, as discussed in \cite{Nikolic2019a, Chin2015}. However, due to modulation instability, the remaining modes will inevitably grow from zero, and ``ruin'' the periodicity of the resultant solution, known as a nonlinear Talbot carpet. To circumvent this issue, we introduce a pruning procedure as follows. After every $x$-evolution step, we ``prune'' the unwanted Fourier modes
\begin{align}
\tilde{\psi}(\omega_j) \rightarrow f(\tilde{\psi}(\omega_j)), \qquad j \neq  M, 2M, \ldots N_t\times M, 
\end{align}
where
\begin{align}
f(\tilde{\psi}(\omega_j))=
	  \tilde{\psi}(\omega_j) \exp\left( -\beta \left|\tilde{\psi}(\omega_j)\right| \right),
\end{align}
for some constant $\beta > 1$.

This pruning procedure was initially introduced in Ref. \cite{Ashour2017} and later refined and used to study nonlinear Talbot carpets in Ref. \cite{Nikolic2019a}.

\begin{figure}
  \sidesubfloat[]{\includegraphics[width=0.9\linewidth]{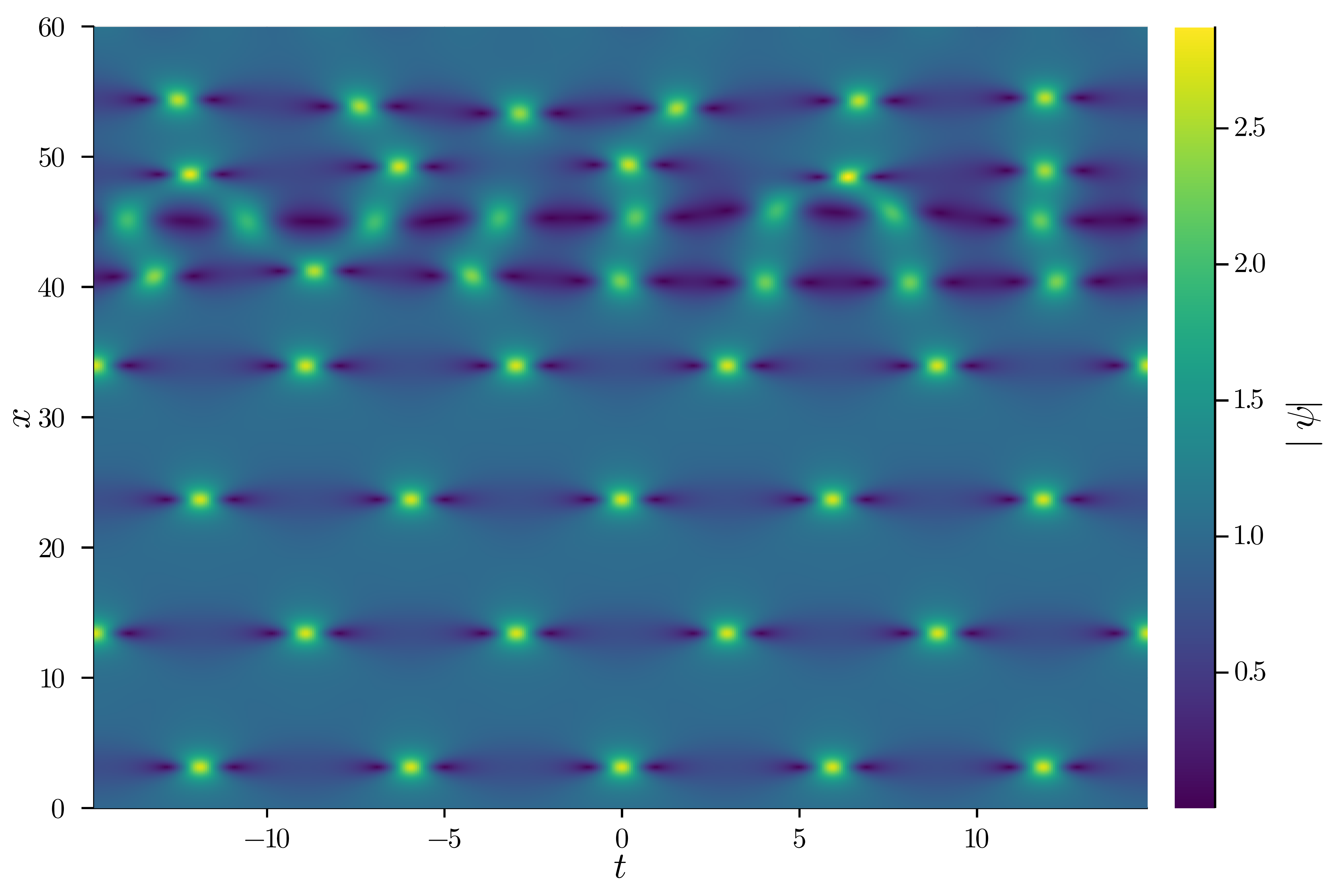}\label{fig:NLTC_psi_noprune}}\\
  \sidesubfloat[]{\includegraphics[width=0.9\linewidth]{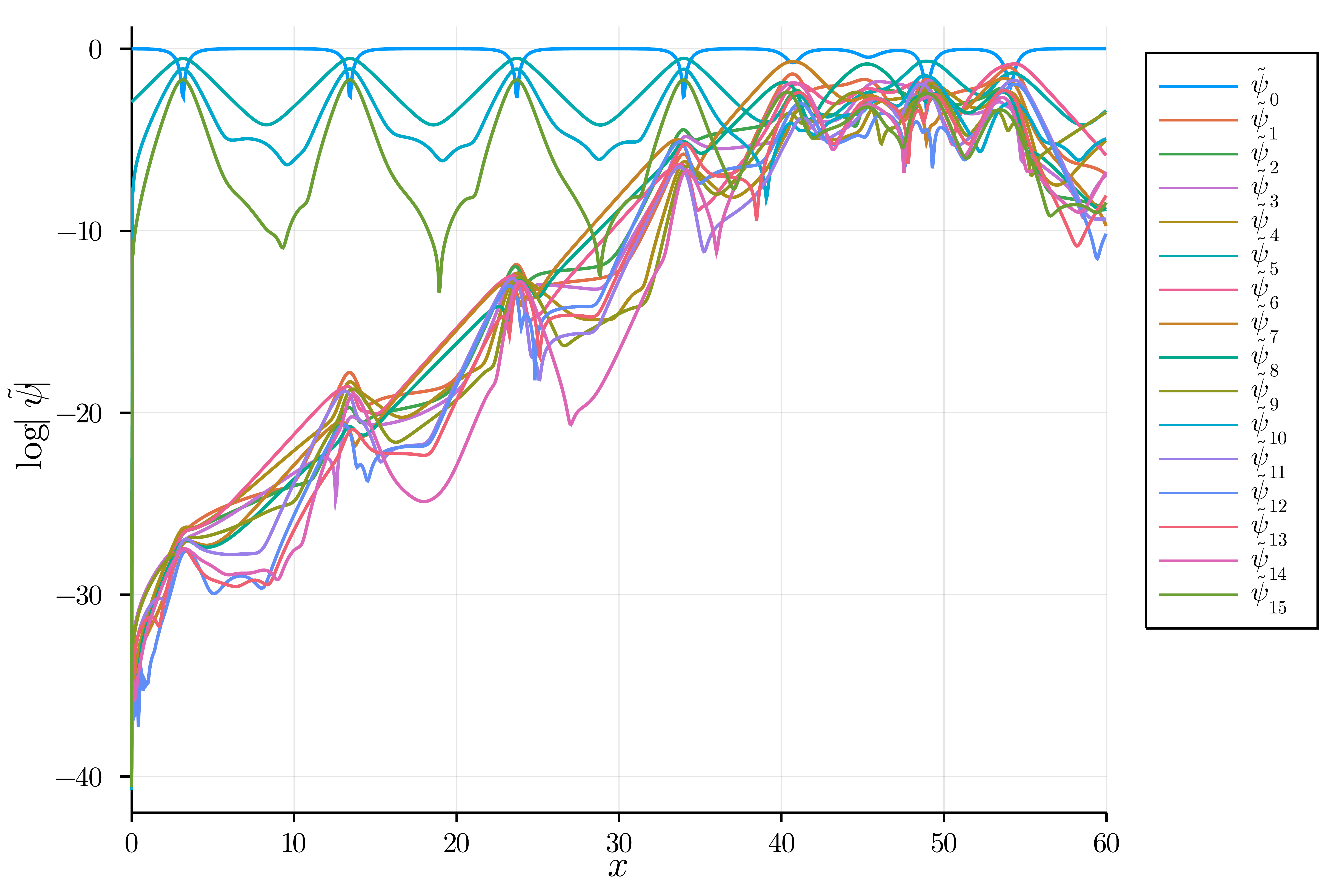}\label{fig:NLTC_spectrum_noprune}}
  \caption{The results of Example \ref{lst:ex3} with pruning disabled ($\beta = 0$) (a) The absolute value of the wave envelope, $|\psi(x, t)|$, showing a nonlinear Talbot carpet destroyed due to modulation instability of the ``non-fundamental'' modes. (b) The spectrum of the solution, showing the growth of the aforementioned modes. \label{fig:no_pruning}}
\end{figure}

\begin{figure}
  \sidesubfloat[]{\includegraphics[width=0.9\linewidth]{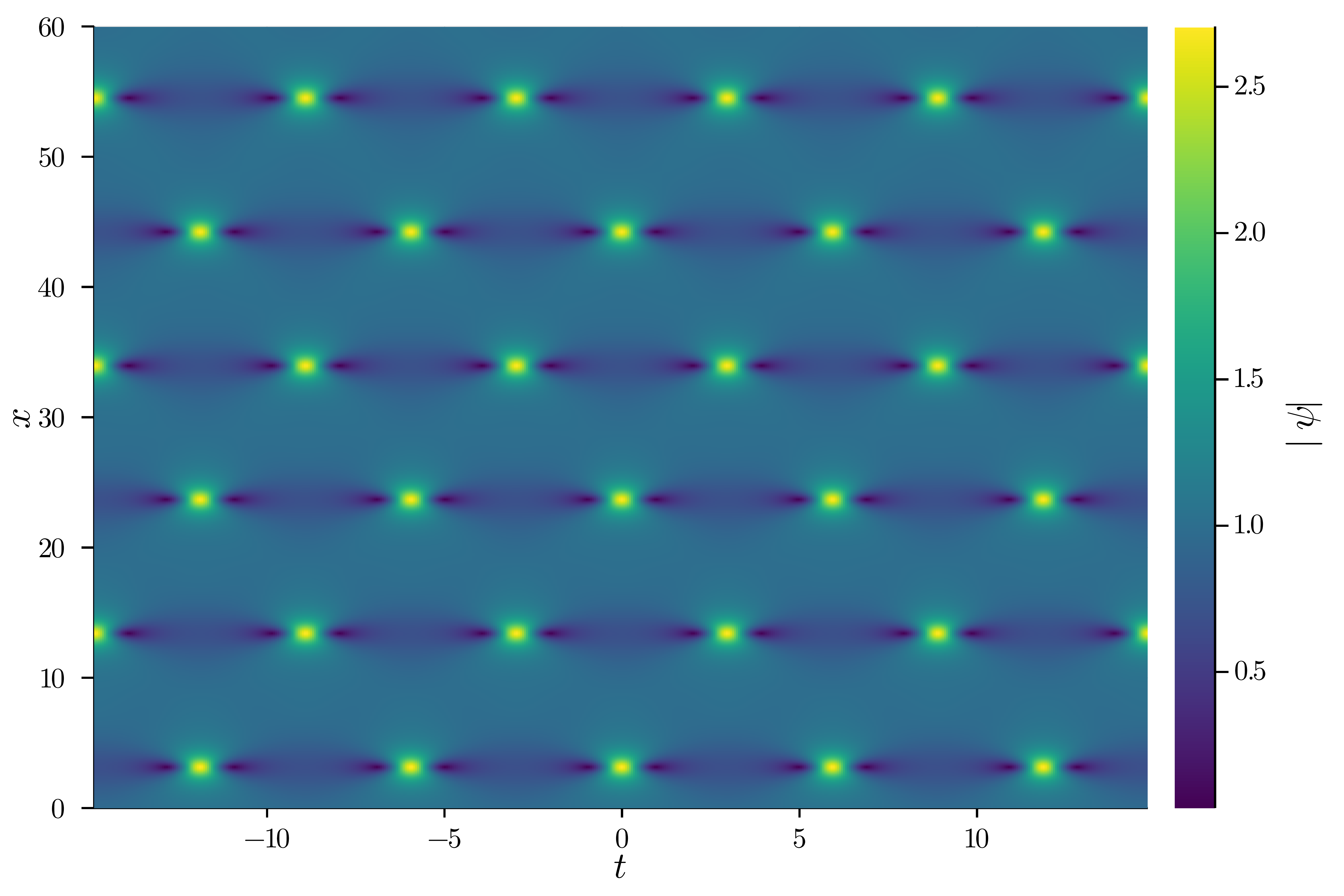}\label{fig:NLTC_psi_prune}}\\
  \sidesubfloat[]{\includegraphics[width=0.9\linewidth]{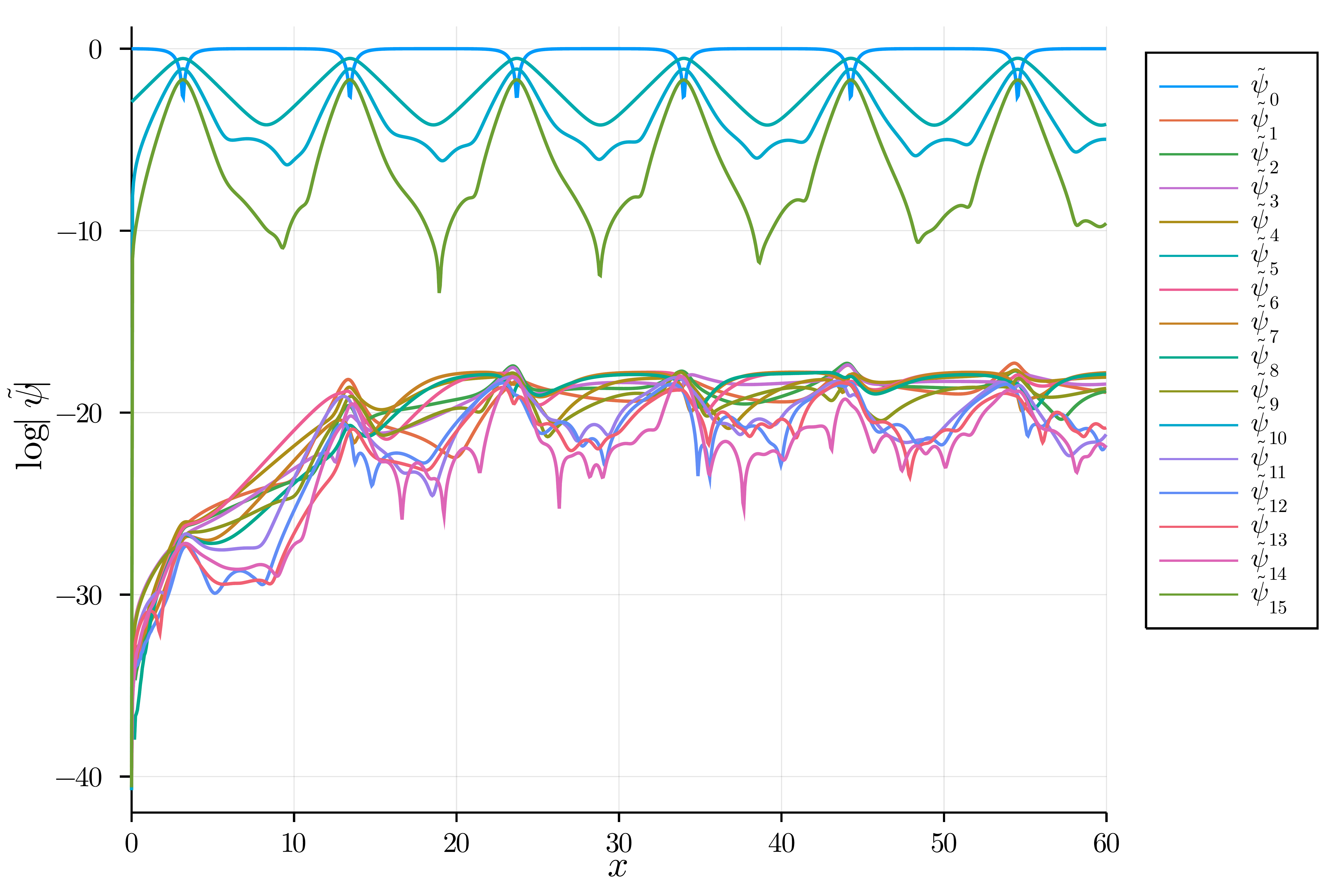}\label{fig:NLTC_spectrum_prune}}
  \caption{The results of Example \ref{lst:ex3}, with pruning enabled ($\beta = 10$).(a) The absolute value of the wave envelope, $|\psi(x, t)|$, showing a perfect nonlinear Talbot carpet. (b) The spectrum of the solution, demonstrating the suppressed ``non-fundamental'' modes. \label{fig:pruning}}
\end{figure}

For example, suppose we use 3 periods. Then, we expect the 3$^\text{rd}$, 6$^\text{th}$, 9$^\text{th}$, etc. Fourier modes to grow together in lockstep. Thus, we can exponentially prune the 1$^\text{st}$, 2$^\text{nd}$, 4$^\text{th}$, 5$^\text{th}$, etc. Fourier modes, as demonstrated above, to stop their spurious growth that ``ruins'' the periodicity of the nonlinear Talbot carpet. See Example \ref{lst:ex3} for a demonstration with 5 periods. The results of this example without ($\beta$ = 0) and with ($\beta = 10$) pruning are shown in Fig. \ref{fig:no_pruning} and Fig. \ref{fig:pruning}, respectively.

\begin{lstlisting}[language=Julia, float, floatplacement=H!, caption={Running a simulation with pruning}, captionpos=b, label={lst:ex3}]
λ, T, Ω = params(a = 0.36)

xᵣ = 0=>60
box = Box(xᵣ, T, dx=1e-4, Nₜ = 512, n_periods = 5)

coeff = [(2.7 + 4.6im)*1e-2]
ψ₀, A₀ = ψ₀_periodic(coeff, box, Ω)

# Set β = 0 to turn off pruning (default behavior when β is not set)
sim = Sim(λ, box, ψ₀, T4A_TJ!, β = 10.0)

solve!(sim)
\end{lstlisting}

\section{Solving the NLSE Analytically \label{sec:analytical}}
The most general extended nonlinear Schr\"odinger equation supported by the Darboux transformation scheme in \texttt{NonlinerSchrodinger.jl} is of the form
\begin{align}
    i{\psi _x} + S[\psi (x,t)] - i\alpha H[\psi (x,t)] + \gamma P[\psi (x,t)] - i\delta Q[\psi (x,t)] = 0,
    \label{eq:extended_NLSE}
\end{align}
where
\begin{align}
\begin{aligned}
S[\psi (x,t)] &= \frac{1}{2}{\psi _{tt}} + {\left| \psi  \right|^2}\psi, \\
H[\psi (x,t)] &= {\psi _{ttt}} + 6{\left| \psi  \right|^2}{\psi _t}, \\
P[\psi (x,t)] &= {\psi _{tttt}} + 8{\left| \psi  \right|^2}{\psi _{tt}} + 6{\left| \psi  \right|^4}\psi + 4{\left| {{\psi _t}} \right|^2}\psi + 6{\psi _t}^2{\psi ^*} + 2{\psi ^2}\psi _{tt}^*, \\
Q[\psi (x,t)] &= {\psi _{ttttt}} + 10{\left| \psi  \right|^2}{\psi _{ttt}} + 30{\left| \psi  \right|^4}{\psi _t} + 10\psi {\psi _t}\psi _{tt}^* + 10\psi \psi _t^*{\psi _{tt}} + 20{\psi ^*}{\psi _t}{\psi _{tt}} + 10\psi _t^2\psi _t^*.
\end{aligned}
\end{align}
Special cases include the cubic nonlinear Schrodinger equation \eqref{eq:nlse} ($\alpha = \gamma = \delta = 0$), the Hirota equation \cite{Hirota1973, Nikolic2017, Ankiewicz2010, Tao2012} ($\alpha \neq 0, \gamma = \delta = 0$) the Lakshmanan-Porsezian-Daniel (LPD) equation \cite{Porsezian1992,Wang2013,Lakshmanan1988} ($\gamma \neq 0, \alpha = \delta = 0$) and the Quintic nonlinear Schrodinger equation (QNLSE) \cite{Nikolic2019,Chowdury2015a,Chowdury2015} ($\delta \neq 0, \alpha = \gamma = 0$). 

For the sake of simplicity, we will restrict ourselves to the cubic nonlinear Schr\"odinger equation \eqref{eq:nlse} in the discussion that follows as a prototypical example, and provide an example of the full extended equation \eqref{eq:extended_NLSE} in Sec. \ref{sec:ex9} (Example \ref{lst:ex9}). The extension of the Darboux transformation (Sec. \ref{sec:DT}) to the extended equation for the soliton (Sec. \ref{sec:seed_0}) and breather (Sec. \ref{sec:seed_exp}) seeds is simple but tedious. It is highly non-trivial for the cnoidal seeds (Sec. \ref{sec:seed_dn} and \ref{sec:seed_cn}, see Ref. \cite{Nikolic2019}) and is not yet implemented in this package.
\subsection{The Lax System \label{sec:lax}}
It is well known that the cubic NLSE \eqref{eq:nlse} can be written as the compatibility condition of the following system \cite{Zakharov1972, Matveev1991, Akhmediev1997, Gu2004}
\begin{align}
\begin{aligned}
R_t &= L R,\\
R_x &= B R,
\end{aligned}
\label{eq:nlse_lax}
\end{align}
where
\begin{align}
L = \begin{pmatrix} -i\lambda & \psi \\ -\psi^* & i \lambda \end{pmatrix},\quad \quad
B = \begin{pmatrix} -i\lambda^2 + \frac{i}{2} |\psi|^2 & \lambda \psi + \frac{i}{2} \psi_x \\ -\lambda \psi^* + \frac{i}{2} \psi_x^* & i\lambda^2 - \frac{i}{2} |\psi|^2  \end{pmatrix}.
\end{align}
$\lambda$ is an isospectral complex eigenvalue, i.e. $\lambda_x = 0$. By compatibility condition we mean that $(R_x)_t = (R_t)_x$ is only satisfied when $\psi$ is a solution of\ \eqref{eq:nlse}, as one can easily check. This condition is also called the zero-curvature condition as it has deep roots in differential geometry \cite{Faddeev2007}. The system \eqref{eq:nlse_lax} is known as the Lax system \cite{Lax1968} of the NLSE. We refer interested readers to \cite{Matveev1991, Gu2004} for more details.
\subsection{The Darboux Transformation \label{sec:DT}}
\begin{figure}
    \centering
    \includegraphics[width=0.9\linewidth]{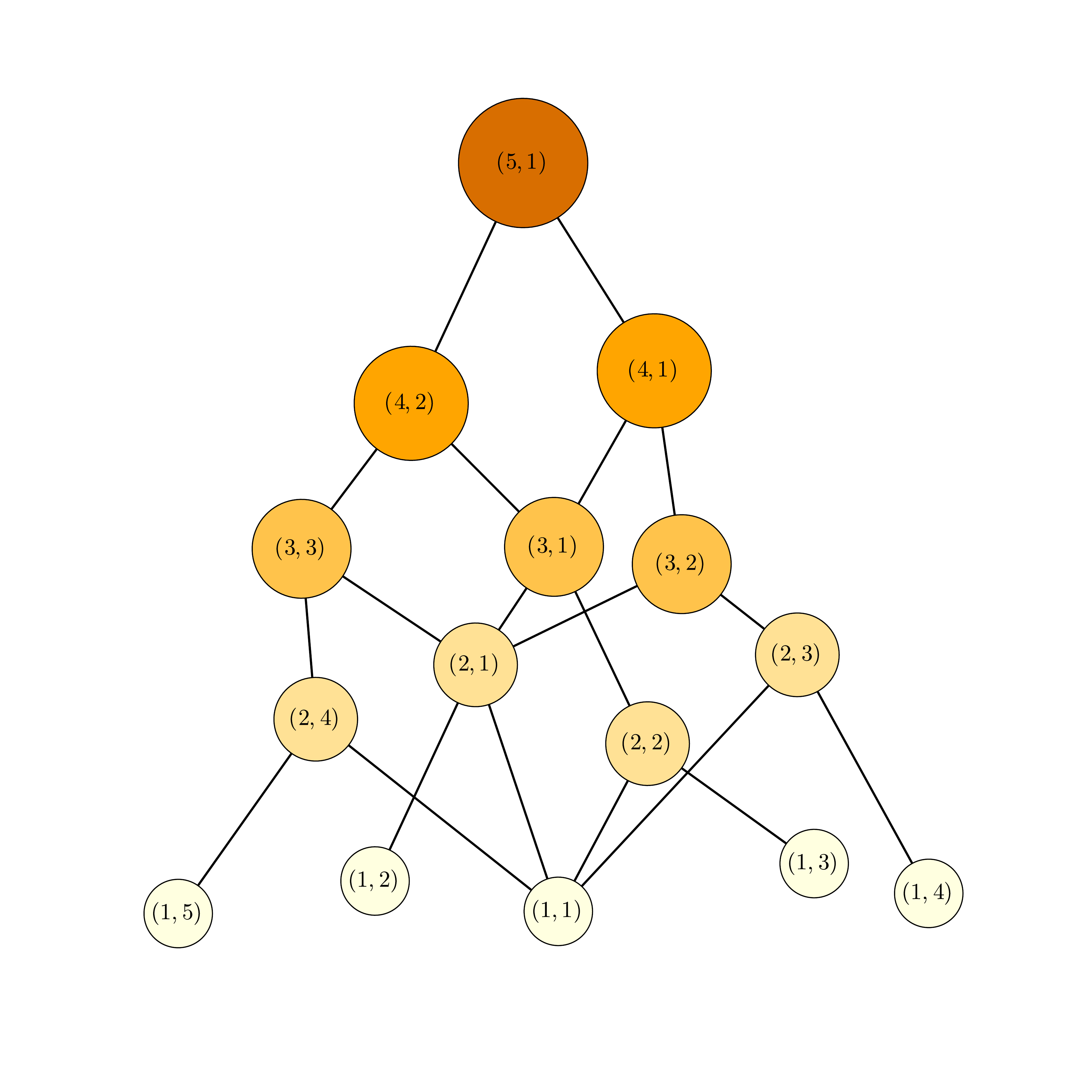}
    \caption{A graphical representation of the Darboux transformation recursion for a fifth order solution. Each ordered pair in the nodes represents a value of $(n,p)$ in equations \eqref{eq:rs_rec}. It can be seen that functions with $p=1$ are used multiple times, i.e., they are connected to multiple nodes in the ``level'' above them. In this package, each function is computed only once and then memoized (i.e., cached), and the cached value is used whenever needed to improve performance dramatically. Graphs similar to this one can be generated automatically using the script in the visualization section of the documentation for arbitrary solution order $N$.}
    \label{fig:recursion}
\end{figure}

A set of solutions of \eqref{eq:nlse_lax} is written as
\begin{align}
    R = \begin{pmatrix} r_{1,p} \\ s_{1,p} \end{pmatrix},
\end{align}
where $p$ labels each unique solution with a \emph{unique} eigenvalue $\lambda_p$. These solutions also depend, in principle, on arbitrary longitudinal and transverse shifts $(x_p, t_p)$, which appear as integration constants. We can obtain a solution of order $n$ recursively via the Darboux transformation \cite{Akhmediev1997}
\begin{align}
    \psi_n = \psi_{n-1} + \frac{2 (\lambda_n^* - \lambda_n) s_{n,1} r_{n,1}^*}{|r_{n,1}|^2 + |s_{n,1}|^2},
    \label{eq:psi_n_dt}
\end{align}
where
\begin{align}
\begin{aligned}
r_{np}&=[(\lambda_{n-1}^*-\lambda_{n-1})s^*_{n-1,1}r_{n-1,1}s_{n-1,p+1}\\
&+(\lambda_{p+n-1}-\lambda_{n-1})|r_{n-1,1}|^2r_{n-1,p+1}\\
 &+(\lambda_{p+n-1}-\lambda^*_{n-1})|s_{n-1,1}|^2r_{n-1,p+1}]/(|r_{n-1,1}|^2+|s_{n-1,1}|^2),\\
s_{np}&=[  (\lambda_{n-1}^*-\lambda_{n-1})s_{n-1,1}r^*_{n-1,1}r_{n-1,p+1}\\
&+(\lambda_{p+n-1}-\lambda_{n-1})|s_{n-1,1}|^2s_{n-1,p+1}\\
&+(\lambda_{p+n-1}-\lambda^*_{n-1})|r_{n-1,1}|^2s_{n-1,p+1}]/(|r_{n-1,1}|^2+|s_{n-1,1}|^2).
\end{aligned}
\label{eq:rs_rec}
\end{align}
The solution $\psi_0$ in \eqref{eq:psi_n_dt} with $n=1$ is called the ``seed'' solution of \eqref{eq:nlse}, and as long as we can solve for the Lax pair generating functions $r_{1,p}$ and $s_{1,p}$ for such a seed solution, we can then obtain higher-order \emph{analytical} solutions of arbitrary order $n$ via the recursive scheme shown above. A visual representation of the recursion is shown in Fig. \ref{fig:recursion}.

In principle, these equations can be implemented in a computer algebra system, such as Mathematica, and used to obtain symbolic expressions for any desired solution. However, such expressions are often too complicated and not particularly insightful, especially beyond the second order. For example, see Eq. (7) in Ref. \cite{Kedziora2012a} for the second-order breather's analytical form. Furthermore, as we will show below, it is not always possible to find $r_{1,p}$ and $s_{1,p}$ analytically for every seed solution of \eqref{eq:nlse}. In this case, the procedure \emph{must} be implemented numerically. In the sections that follow, we discuss the four possible seed solutions of \eqref{eq:nlse} and how the Lax system \eqref{eq:nlse_lax} is solved for each of them. 
\subsection{The Four Seeds \label{sec:4_seeds}}
In this section, we discuss the four most prominent seed solutions of the NLSE implemented in \texttt{NonlinearSchrodinger.jl} and the higher-order solutions to which they correspond.
\subsubsection{\texorpdfstring{$\psi_0 = 0$}{psi0=0}\label{sec:seed_0}}
The seed solution $\psi_0 = 0$ generates solitons via the Darboux transformation. Plugging $\psi = 0$ into \eqref{eq:nlse_lax}, we get a simple form for $r_{1,p}$ and $s_{1,p}$
\begin{align}
    r_{1,p}(x,t) &= \exp\left(+i \left[\lambda_p (t - t_p) + \lambda_p^2 (x - x_p) - \pi/4 \right] \right), \\
    s_{1,p}(x,t) &= \exp\left(-i \left[\lambda_p (t - t_p) + \lambda_p^2 (x - x_p) - \pi/4 \right] \right),
\end{align}
where $\lambda_p$ are \emph{unique} eigenvalues and $(t_p, x_p)$ are arbitrary shifts in the $t$ and $x$ directions. The phase factor of $\pi/4$ is selected to center the solutions at the origin \cite{Chin2017}.
\subsubsection{\texorpdfstring{$\psi_0 = e^{ix}$}{psi0 = exp(ix)} \label{sec:seed_exp}}
The second seed solution, $\psi_0 = e^{ix}$, corresponds to breathers when passed through the Darboux transformation. Similar to the previous section, we plug $\psi = e^{ix}$ into \eqref{eq:nlse_lax}, and solve the resultant coupled differential equations. We can get a simple closed-form for the Lax pair generating functions 
\begin{align}
    \begin{aligned}
    r_{1,p}(x,t) &= 2i e^{-ix/2}\sin\left(+\chi_p + \frac{1}{2}\Omega_p (t - t_p) +\frac{1}{2}\Omega_p\lambda_p(x-x_p) - \frac{\pi}{4}\right),\\
    s_{1,p}(x,t) &= 2\, e^{+ix/2}\cos\left(-\chi_p+ \frac{1}{2}\Omega_p (t - t_p) + \frac{1}{2}\Omega_p\lambda_p(x-x_p) - \frac{\pi}{4}\right).\\
    \end{aligned}
\end{align}
Here, $\Omega_p = 2 \sqrt{1 + \lambda_p^2}$ is the frequency of the $p^\text{th}$ constituent solution, and $\chi_p = \arccos(\Omega_p/2)/2$. 

\subsubsection{\texorpdfstring{$\psi_0 = \text{dn}(t, m)e^{ix(1-m/2)}$}{psi0 = dn(t,m) exp(i x (1 - m/2))}\label{sec:seed_dn}}

The seed solution $\psi_0 = \text{dn}(t, m)e^{ix(1-m/2)}$ corresponds to breathers on an elliptic $\dn$ background \cite{Kedziora2014}. $m$ is the elliptic parameter (not to be confused with the elliptic modulus $k = \sqrt{m}$) and $\dn$ is one of the Jacobi elliptic functions (JEFs). For a reference on JEFs, see \cite{Schwalm2015}.

In this case, we use the ans\"atze
\begin{align}
\begin{aligned}
r_{1p}(x,t) &= a_{1p}(x,t)e^{\frac{ix}{4}\left(m-2\right)},\\
s_{1p}(x,t) &= b_{1p}(x,t)e^{-\frac{ix}{4}\left(m-2\right)}.
\end{aligned}
\end{align}
By substituting into \eqref{eq:nlse_lax}, and suppressing the subscripts and dependence on $x$ and $t$ for clarity, we get \cite{Kedziora2014}
\begin{align}
\begin{aligned}
a_t &= i\lambda a + ib \dn(t,m), \\
b_t &= -i\lambda b + ia\dn(t,m),\\
a_x &= \half ia \left(2\lambda^2 + m\left(\sn^2(t,m) - \half\right)\right) + b\left(i\lambda\dn(t,m) - \frac{m}{2}\sn(t,m)\cn(t,m)\right), \\
b_x &= -\half ib \left(2\lambda^2 + m\left(\sn^2(t,m) - \half\right)\right) + a\left(i\lambda\dn(t,m) + \frac{m}{2}\sn(t,m)\cn(t,m)\right).
\end{aligned}
\label{eq:abdn}
\end{align}
However, unlike the soliton and breather seeds, these coupled differential equations have no analytical solutions that we know of. We can solve for the profiles and derivatives at $t=0$
\begin{align}
\begin{aligned}
a_{1p}|_{t=0} &= Ae^{i(\chi_p+\frac{1}{2}\Omega_p\lambda_p(x-x_p))} - Be^{-i(\chi_p+\frac{1}{2}\Omega_p\lambda_p(x-x_p))},\\
b_{1p}|_{t=0} &= A e^{i(-\chi_p+\frac{1}{2}\Omega_p\lambda_p(x-x_p))} + Be^{-i(-\chi_p+\frac{1}{2}\Omega_p\lambda_p(x-x_p))},\\
a_{1p,t}|_{t=0} &= i(\lambda_p a_{1p}|_{t=0} + b_{1p}|_{t=0}),\\
b_{1p,t}|_{t=0} &= -i(\lambda_p b_{1p}|_{t=0} - a_{1p}|_{t=0}).
\end{aligned}
\label{eq:profiles_dn}
\end{align}
Here, we have defined $\Omega _p= 2\sqrt{\left(\lambda_p-\frac{m}{4 \lambda_p}\right)^2+1}$ as the frequency of the solution, and  $\chi_p = \arccos\left(\frac{\Omega_p}{2}\right)/2$. Further, we set the integration constants  $A = e^{-i\pi/4} = B^*$ as before to center the solutions at the origin (when the shifts are set to zero). Finally, we get a set of coupled equations along with their initial conditions
\begin{align}
\begin{aligned}
(a_{1p})_t &= i\lambda a(x,t) + ib(x,t) \dn(t,m), \\
(b_{1p})_t &= -i\lambda b(x,t) + ia(x,t)\dn(t,m),\\
a_{1p}|_{t=0} &= 2 i \sin\left(\chi_p+\frac{1}{2}\Omega_p\lambda_p(x-x_p) - \frac{\pi}{4}\right),\\
b_{1p}|_{t=0} &= 2 \cos\left(-\chi_p+\frac{1}{2}\Omega_p\lambda_p(x-x_p) - \frac{\pi}{4}\right).\\
\end{aligned}
\label{eq:abprofiles}
\end{align}

We have ignored the shifts $t_p$ along the temporal direction in this procedure for simplicity. These coupled differential equations must be evolved numerically in $t$ to obtain the final solution. We are not aware of any methods to solve them analytically.

\subsubsection{\texorpdfstring{$\psi_0 = \sqrt{m}\text{cn}(t, m)e^{ix(m - 1/2)}$ }{psi0 = sqrt(m) cn(t,m) exp(i x (m-1/2))} \label{sec:seed_cn}}
The seed $\psi_0 = \sqrt{m}\text{cn}(t, m)e^{ix(m - 1/2)}$ corresponds to solitons on a cnoidal background \cite{Kedziora2014}. The solution process is very similar to what was presented in the previous section for the dnoidal seed. First, we take the ans\"atze
\begin{align}
\begin{aligned}
r_{1p}(x,t) &= a_{1p}(x,t)e^{\frac{-ix}{4}\left(2m - 1\right)},\\
s_{1p}(x,t) &= b_{1p}(x,t)e^{\frac{ix}{4}\left(2m -1\right)}.
\end{aligned}
\end{align}
Using these ans\"atze and the process previously outlined, we obtain \cite{Kedziora2014}
\begin{align}
\begin{aligned}
(a_{1p})_t &= i\lambda a(x,t) + i \sqrt{m} b(x,t) \cn(t,m), \\
(b_{1p})_t &= -i\lambda b(x,t) + i \sqrt{m} a(x,t)\cn(t,m),\\
a_{1p}|_{t=0} &= 2 i \sin\left(\chi_p+\frac{1}{2}\Omega_p\lambda_p(x-x_p) - \frac{\pi}{4}\right),\\
b_{1p}|_{t=0} &= 2 \cos\left(-\chi_p+\frac{1}{2}\Omega_p\lambda_p(x-x_p) - \frac{\pi}{4}\right).\\
\end{aligned}
\label{eq:profiles_cn}
\end{align}
Where now $\Omega_p = 2 \sqrt{m} \sqrt{1 + \frac{1}{m}\left(\lambda_p - \frac{1}{4\lambda_p}\right)^2}$ and $\chi_p = \half \arccos\left(\frac{\Omega_p}{2\sqrt{m}}\right)$.

\subsection{Maximal Intensity Families \label{sec:maximal_intensity}}

For the periodic seeds discussed in Sec. \ref{sec:seed_exp} and \ref{sec:seed_dn}, there are so-called maximal intensity families that match the constituent breathers' periods to each other. See \cite{Chin2016} for the uniform background case and \cite{Ashour2018} for the dnoidal background case. 

In the uniform background case, it is simple to match these periods to each other: $\Omega_p = p \Omega$, where $\Omega \equiv \Omega_1$ is the period of the first-order constituent breather. This leads to the following equations for the imaginary parts of the eigenvalues
\begin{align}
    \nu_p = \sqrt{p^2 (\nu^2 - 1) + 1},
    \label{eq:nu_p_maximal}
\end{align}
where $\nu_p = \mathfrak{Im}(\lambda_p)$ and $\nu \equiv \nu_1 = \mathfrak{Im}(\lambda_1)$ is the so-called fundamental eigenvalue of the maximal intensity family. Note that, for an $N^\text{th}$ order breather, we must have $\nu > \nu^*$ for all the $\nu_p$ to be real, where
\begin{align}
    \nu^* = \sqrt{1 - \frac{1}{N^2}}.
\end{align}
Effectively, this collapses the parameter space of an $N^\text{th}$ order breather from $N$ dimensions to $1$ dimension, ignoring the spatiotemporal shifts.

For the breathers on the dnoidal background, the process is similar but more involved \cite{Ashour2018}. We start by defining the following function
\begin{align}
    G_p(m, \nu) = m^2 p^2 + 8 (m - 2) (p^2 - 1) \nu^2 + 16 p^2 \nu^4 .
    \label{eq:nu_p_maximal_dn}
\end{align}
Now, we get
\begin{align}
    \nu_p = \frac{\sqrt{G_p(m,\nu) + \sqrt{[G_p(m, \nu)]^2 - 64 m^2 \nu^4}}}{4 \sqrt{2} \nu},
\end{align}
One can verify \eqref{eq:nu_p_maximal_dn} reduces to \eqref{eq:nu_p_maximal} when $m = 0$. There is an analogous but more complicated expression for $\nu^*$ given in Ref. \cite{Ashour2018}. 

However, one must also match the breathers to the background for a truly periodic solution. In this case, $T_B = q T_{\text{dn}}$ where $q$ is a positive integer, $T_B$ is the period of the fundamental breather characterized by $\nu$ and $T_{\dn}$ is the period of the dnoidal background, characterized by $m$.

Skipping the details of the derivation, the implementation in \texttt{NonlinearSchrodinger.jl} uses the following equation to compute $\nu$ given a value of $m$
\begin{align}
    \nu &= \frac{1}{2} \sqrt{2 - F^2 - m + 2 \sqrt{(F^2 - 1)(F^2-1 + m)}},\\
    F &= \frac{\pi}{2 q K(m)}.
\end{align}
Here, $K(m)$ is the complete elliptic integral of the first kind \cite{Schwalm2015}. We are not aware of a method to invert this equation analytically to obtain $m$ given a value of $\nu$.


These maximal intensity families are implemented in \texttt{NonlinearSchrodinger.jl} via the functions $\lambda$\texttt{\_maximal} (to get a set of $\lambda_p$ given $\lambda$) and $\lambda$\texttt{\_given\_m} (to get a value of $\lambda$ given $m$ and an integer $q$). The code automatically checks that the provided $\lambda$ is large enough (i.e., $\nu > \nu^*$) for a given breather order $N$ and returns an error otherwise. For an example on how to use these functions, see Sec. \ref{sec:ex5} (Example \ref{lst:ex5}) and \ref{sec:ex7} (Example \ref{lst:ex7}).

\subsection{Numerical Implementation}
In \texttt{NonlinearSchrodinger.jl}, we implement the Darboux transformation numerically and use it to study these analytical solutions. If the seed solution leads to exact expressions for $r_{1,p}$ and $s_{1,p}$, then the equations are implemented as is, and thus they are accurate within the limits of double-precision (i.e. $\psi_0 = 0$ (Sec. \ref{sec:seed_0}) and $\psi_0 = e^{ix}$ (Sec. \ref{sec:seed_exp}). If this is not possible (i.e. the cnoidal and dnoidal seeds discussed in Sec. \ref{sec:seed_dn} and \ref{sec:seed_cn}), then we obtain $r_{1,p}$ and $s_{1,p}$ using standard numerical algorithms \footnote{Currently, the cnoidal and dnoidal seeds assume that the solution is symmetric about $t=0$. This constraint will be lifted in a future version.}. We specifically use Tsitouras' 5/4 Runge-Kutta method \cite{Tsitouras2011} as implemented in Julia's \texttt{DifferentialEquations.jl} library \cite{Rackauckas2017}.

\subsection{Using \texttt{NonlinearSchrodinger.jl} for the Darboux Transformation \label{sec:using_NLSS_DT}}
In what follows, we present several examples demonstrating how to use the package to perform Darboux transformation calculations using all four seeds.
\subsubsection{Example 4: Seven-Soliton Collision \label{sec:ex4}}
\begin{figure}
    \centering
    \includegraphics[width=0.96\linewidth]{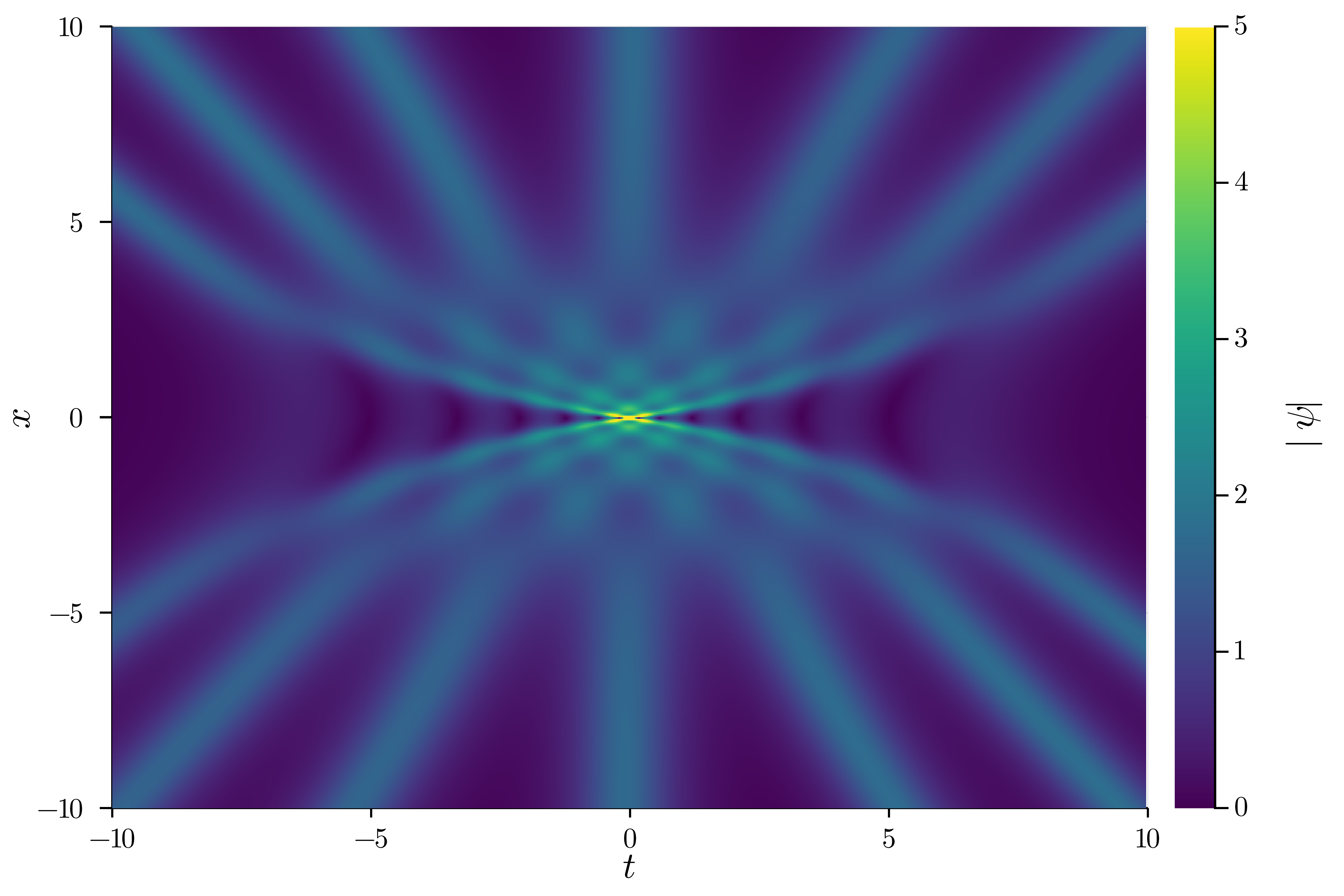}
    \caption{A seven-soliton collision obtained using the Darboux transformation as demonstrated in Example \ref{lst:ex4}. The color scale is cut off at $|\psi| = 5$ to better highlight the low-intensity features of the collision. The actual peak-height of the soliton at $(x,t) = (0,0)$ is 11.9.} 
    \label{fig:7soliton}
\end{figure}

In this example, we demonstrate how to obtain a highly complicated solution, a seven-soliton collision. The first step is to create a calculation ``box'', as was done in the simulation examples. In this case, it is often easier to specify the number of $N_x$ grid points directly instead of $dx$.
\begin{lstlisting}[language=Julia]
xᵣ = -10=>10
T = 20
box = Box(xᵣ, T, Nₓ=1000, Nₜ = 1000)
\end{lstlisting}
We can use any box size since there is no need to worry about boundary conditions in Darboux transformation calculations. The next step is deciding on the eigenvalues $\lambda$ and the spatiotemporal shifts. In this example, since we are interested in a seven-soliton solution, we have seven of each of these parameters. Recall that the eigenvalues must be unique in the Darboux transformation scheme. We use complex eigenvalues to give the solitons a velocity (or ``tilt'') in the $tx$-plane, determined by the eigenvalue's real part. Further, we set all the shifts to zero so that the solitons collide at the origin.
\begin{lstlisting}[language=Julia]
λ = [-0.45 + 0.775im, -0.35 + 0.8im, -0.25 + 0.825im, 0.85im, 
     0.25 + 0.875im, 0.35 + 0.9im, 0.45 + 0.925im]
xₛ = [0.0, 0.0, 0.0, 0.0, 0.0, 0.0, 0.0]
tₛ = [0.0, 0.0, 0.0, 0.0, 0.0, 0.0, 0.0]
\end{lstlisting}
The size of all three arrays must always be equal and determines the order of the solution.
Finally, we create the calculation object and solve it.
\begin{lstlisting}[language=Julia]
seed = "0"
calc = Calc(λ, tₛ, xₛ, seed, box)

solve!(calc)
\end{lstlisting}
We can compute the integrals of motion as shown before using the \texttt{compute\_IoM!} function. The spectrum is calculated automatically as with simulations. The final calculation results can be accessed and plotted in the same way shown in Sec. \ref{sec:ex1}. This example in its entirety is shown in code listing \ref{lst:ex4}, and the result is depicted in Fig. \ref{fig:7soliton}.

\begin{lstlisting}[language=Julia, float, floatplacement=H!, caption={Seven-soliton collision computed via the Darboux transformation}, captionpos=b, label={lst:ex4}]
xᵣ = -10=>10
T = 20
box = Box(xᵣ, T, Nₓ=1000, Nₜ = 1024)

λ = [-0.45 + 0.775im, -0.35 + 0.8im, -0.25 + 0.825im, 0.85im, 
     0.25 + 0.875im, 0.35 + 0.9im, 0.45 + 0.925im]
xₛ = [0.0, 0.0, 0.0, 0.0, 0.0, 0.0, 0.0]
tₛ = [0.0, 0.0, 0.0, 0.0, 0.0, 0.0, 0.0]

seed = "0"
calc = Calc(λ, tₛ, xₛ, seed, box) 

solve!(calc)
\end{lstlisting}

\subsubsection{Example 5: Fifth-Order Breather on a Uniform Background \label{sec:ex5}}

The purpose of this example is to demonstrate how to deal with breathers, where one often wants the transverse box size to be a multiple of their period. Furthermore, we demonstrate the usage of the function $\lambda$\texttt{\_maximal} to generate a maximal intensity (i.e., fully periodic) solution.

First, we start by creating a box for the calculation, just as before. However, in this case, we would like the box size to be three periods of the breather. We utilize the function \texttt{params} to compute the period of the breather, and use the \texttt{n\_periods} argument of the \texttt{Box} constructor to get three periods.
\begin{lstlisting}[language=Julia]
xᵣ = -10=>10
λ₁ = 0.98im # The fundamental eigenvalue
λ, T, Ω = params(λ = λ₁)
box = Box(xᵣ, T, Nₓ=1000, Nₜ = 1024, n_periods = 3)
\end{lstlisting}

The next step is to set up the eigenvalues and shifts. Instead of specifying the eigenvalues explicitly as in the previous example, we employ the $\lambda$\texttt{\_maximal} function to generate them automatically, with $N=5$ to get a fifth-order solution. These eigenvalues guarantee its periodicity, as discussed in Sec. \ref{sec:maximal_intensity}.

\begin{figure}
    \centering
    \includegraphics[width=0.96\linewidth]{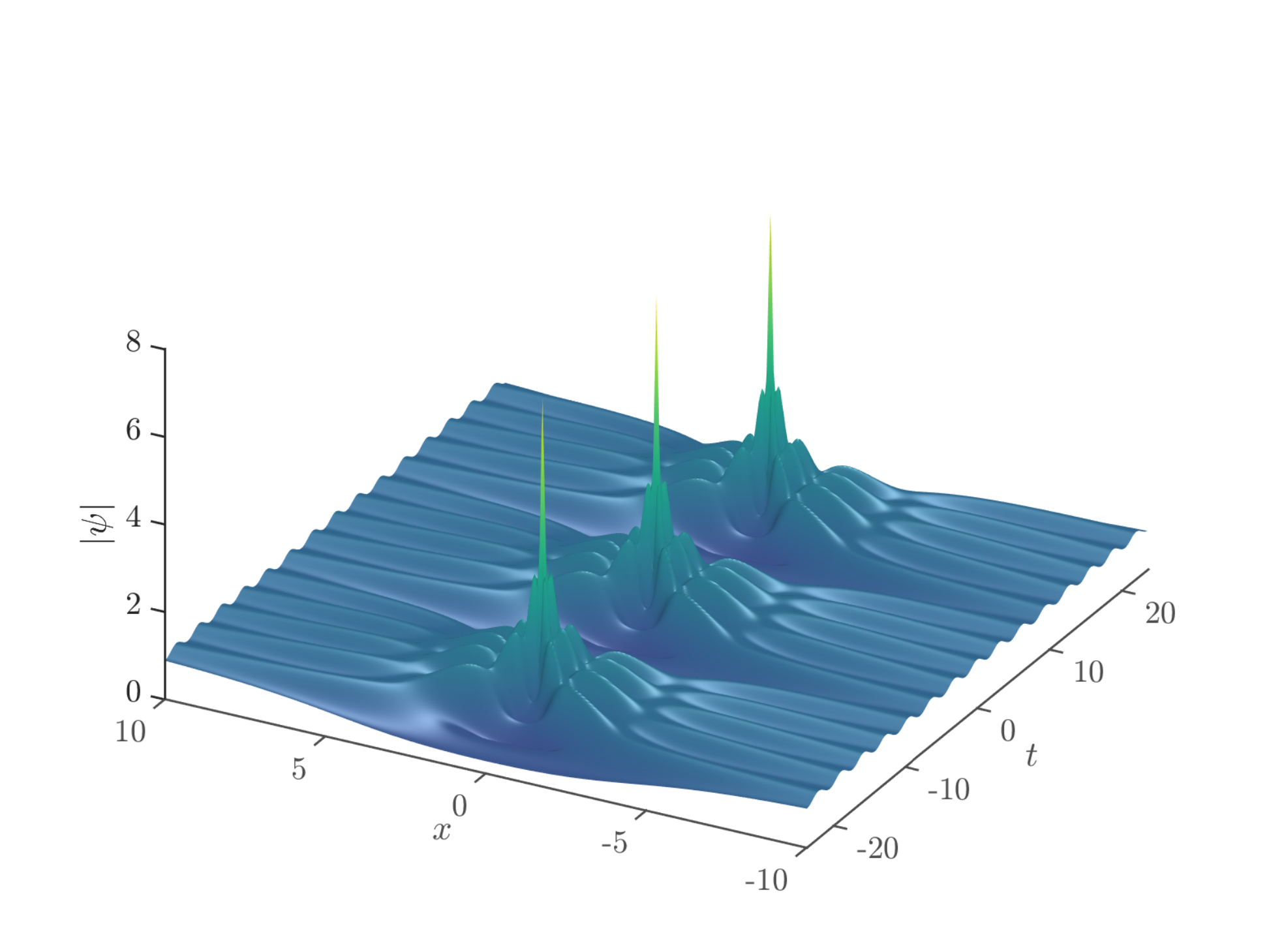}
    \caption{The result of Example \ref{lst:ex5}, a fully periodic fifth-order breather.}
    \label{fig:5AB}
\end{figure}

\begin{lstlisting}[language=Julia]
λ = λ_maximal(λ₁, 5) # array of 5 eigenvalues
xₛ = [0.0, 0.0, 0.0, 0.0, 0.0]
tₛ = [0.0, 0.0, 0.0, 0.0, 0.0]
\end{lstlisting}
This results in the following values of $\lambda$:
\begin{lstlisting}[language=Julia]
λ = [0.98im, 0.9173875952943771im, 0.8022468448052632im, 
     0.6053098380168617im, 0.0999999999999974im]
\end{lstlisting}

Finally, just as before, we define the \texttt{Calc} structure and solve it. The full example is shown in code listing \ref{lst:ex5}, and the result of the calculation is depicted in Fig. \ref{fig:5AB}, highlighting the full periodicity of the solution. As a pedagogical exercise, we urge users to change these automatically generated eigenvalues by hand, run the calculation and observe the differences.

\begin{lstlisting}[language=Julia, float, floatplacement=H!, caption={Calculating a maximal intensity fifth order breather via the Darboux transformation}, captionpos=b, label={lst:ex5}]
xᵣ = -10=>10
λ₁ = 0.98im
λ, T, Ω = params(λ = λ₁)
box = Box(xᵣ, T, Nₓ=1000, Nₜ = 1024, n_periods = 3)

λ = λ_maximal(λ₁, 5)
xₛ = [0.0, 0.0, 0.0, 0.0, 0.0]
tₛ = [0.0, 0.0, 0.0, 0.0, 0.0]

seed = "exp"
calc = Calc(λ, tₛ, xₛ, seed, box) 

solve!(calc)
\end{lstlisting}

\subsubsection{Example 6: Three-Soliton Collision a Cnoidal Background \label{sec:ex6}}
\begin{figure}
    \centering
    \includegraphics[width=0.96\linewidth]{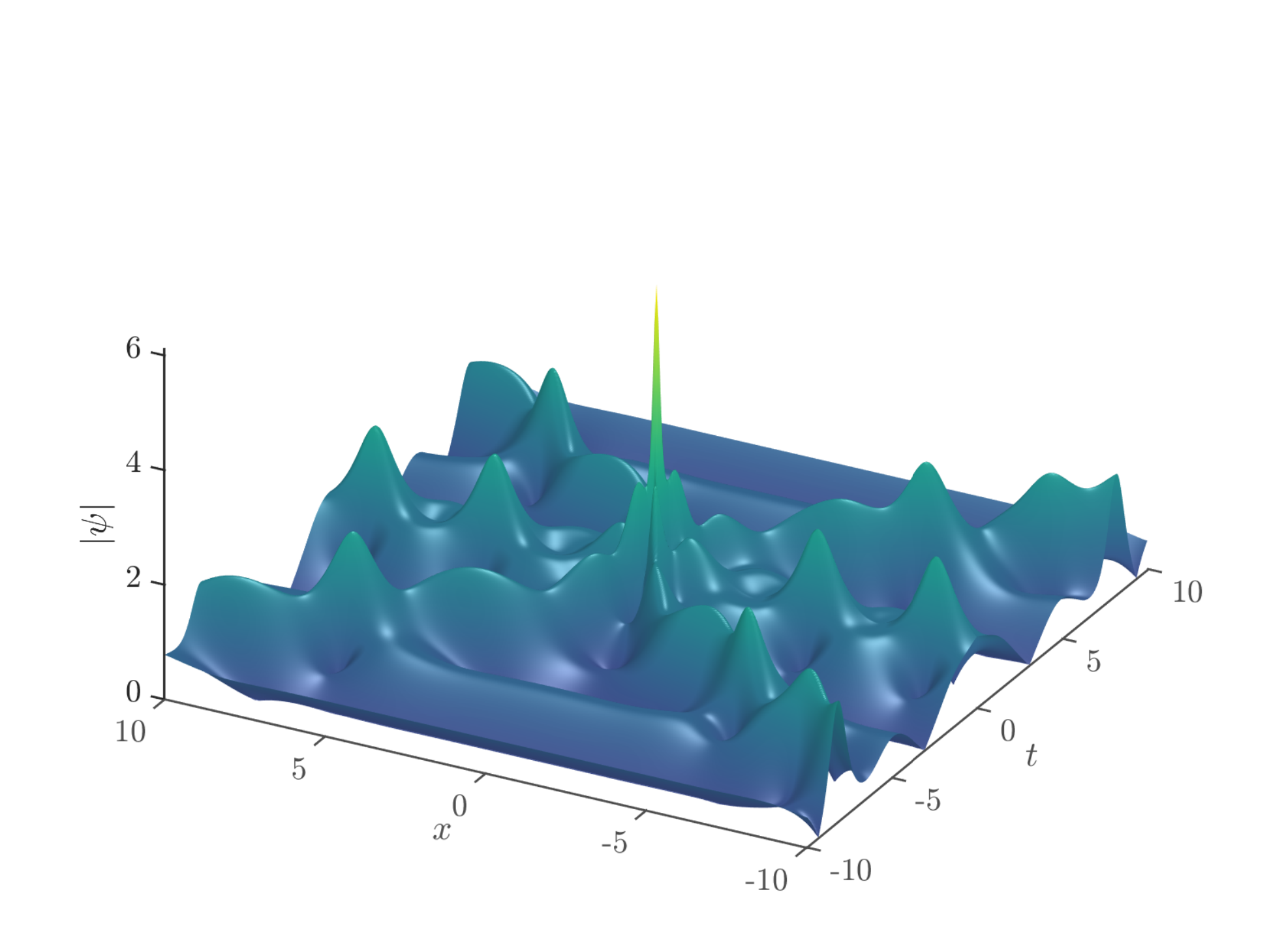}
    \caption{Results of Example \ref{lst:ex6}, showcasing the collision of three solitons on a cnoidal background with $m=1/2$.}
    \label{fig:3_soliton_cnoidal}
\end{figure}

This example is quite simple and is analogous to Example \ref{lst:ex4} (Section \ref{sec:ex4}), but on a cnoidal background. The main point to note here is that $N_t$ used in the box is half the number of nodes used in the integration of Eq. \eqref{eq:profiles_cn}. Thus, $N_t$ should be treated as a convergence parameter. The full example is shown in code listing \ref{lst:ex6}, and the result is depicted in Fig. \ref{fig:3_soliton_cnoidal}.

\begin{lstlisting}[language=Julia, float, floatplacement=H!, caption={Collision of three solitons on a cnoidal background}, captionpos=b, label={lst:ex6}]
xᵣ = -10=>10
T = 20

box = Box(xᵣ, T, Nₓ=1000, Nₜ = 1024)
λ = [-0.3+0.85im, 0.9im, 0.3+0.95im]
xₛ = [0.0, 0.0, 0.0]
tₛ = [0.0, 0.0, 0.0]

seed = "cn"
calc = Calc(λ, tₛ, xₛ, seed, box, m = 0.5) 

solve!(calc)
\end{lstlisting}

\subsubsection{Example 7: First-Order Breather on a Dnoidal Background \label{sec:ex7}}

The purpose of this example is to demonstrate the usage of the function $\lambda$\texttt{\_given\_m} to generate a maximal intensity breather on a dnoidal background. As explained in Sec. \ref{sec:maximal_intensity}, one picks a value of $m$, together with an integer $q \geq 2$ to compute a fundamental eigenvalue $\lambda$. This $\lambda$ can then be used to compute a higher-order breather's eigenvalues as before if required. In this example, we restrict ourselves to a first-order breather to best highlight the effect of background matching. 

\begin{figure}
    \centering
    \includegraphics[width=0.96\linewidth]{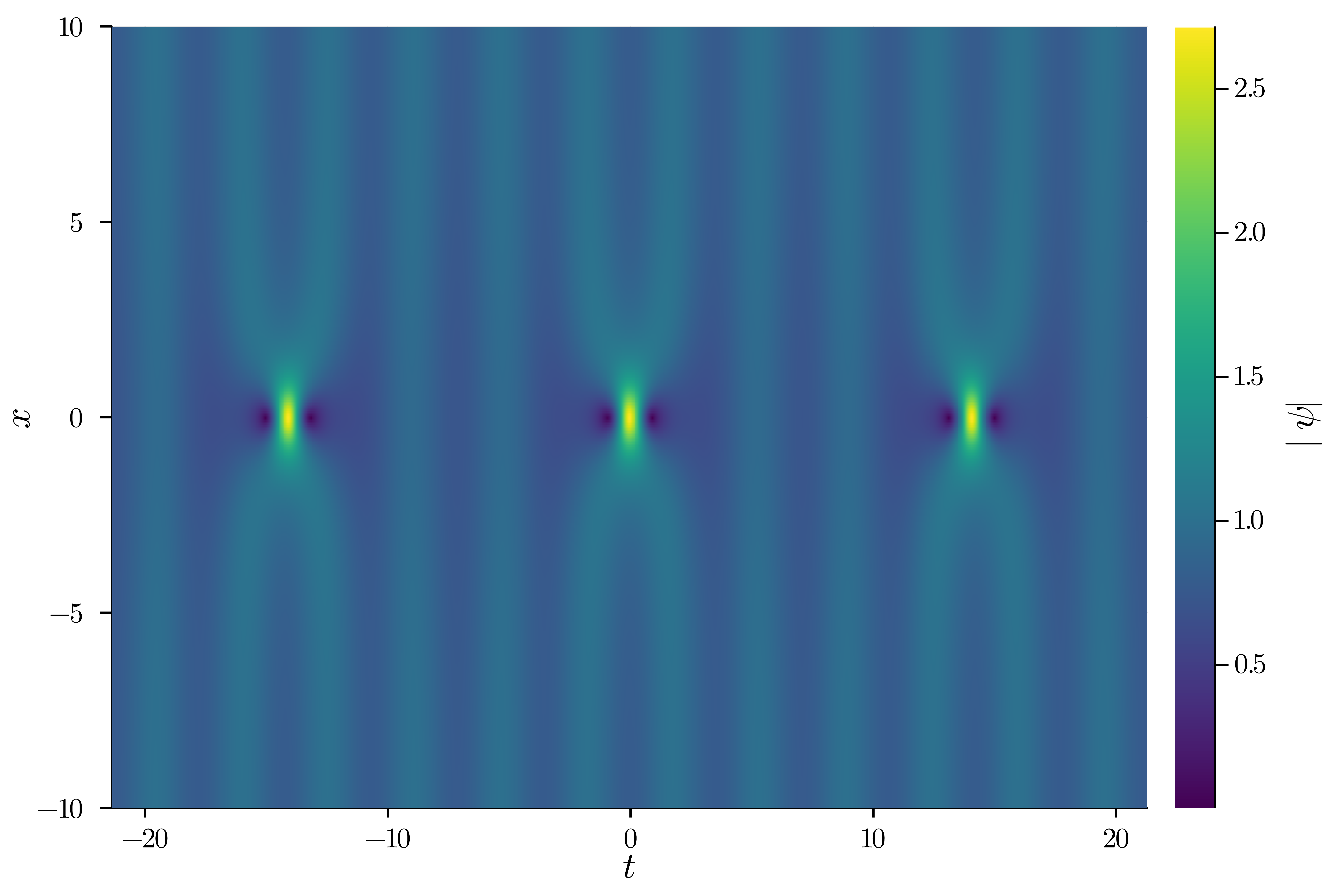}
    \caption{First-order breather fully matched to the underlying dnoidal background, as computed in Example \ref{lst:ex7}.}
    \label{fig:breather_dn}
\end{figure}

This section should be self-explanatory with the background previously given. The full example is shown in code listing \ref{lst:ex7} and depicted in Fig. \ref{fig:breather_dn}. As before, we urge users to change the value of $\lambda$ automatically computed by $\lambda$\texttt{\_given\_m} to see the effect on the solution.

\begin{lstlisting}[language=Julia, float, floatplacement=H!, caption={First-order breather on a dnoidal background with $q=4$ and $m=2/5$.}, captionpos=b, label={lst:ex7}]
m = 2/5
λ = λ_given_m(m, q=4)
λ, T, Ω = params(λ = λ, m=m)
xᵣ = -10=>10
box = Box(xᵣ, T, Nₓ=1000, Nₜ = 1024, n_periods = 3)

xₛ = [0.0]
tₛ = [0.0]

seed = "dn"
calc = Calc(λ, tₛ, xₛ, seed, box, m=m) 

solve!(calc)
\end{lstlisting}

\subsubsection{Example 8: Combining Darboux Transformations and Simulations \label{sec:ex8}}

\begin{figure}
    \centering
    \includegraphics[width=0.9\linewidth]{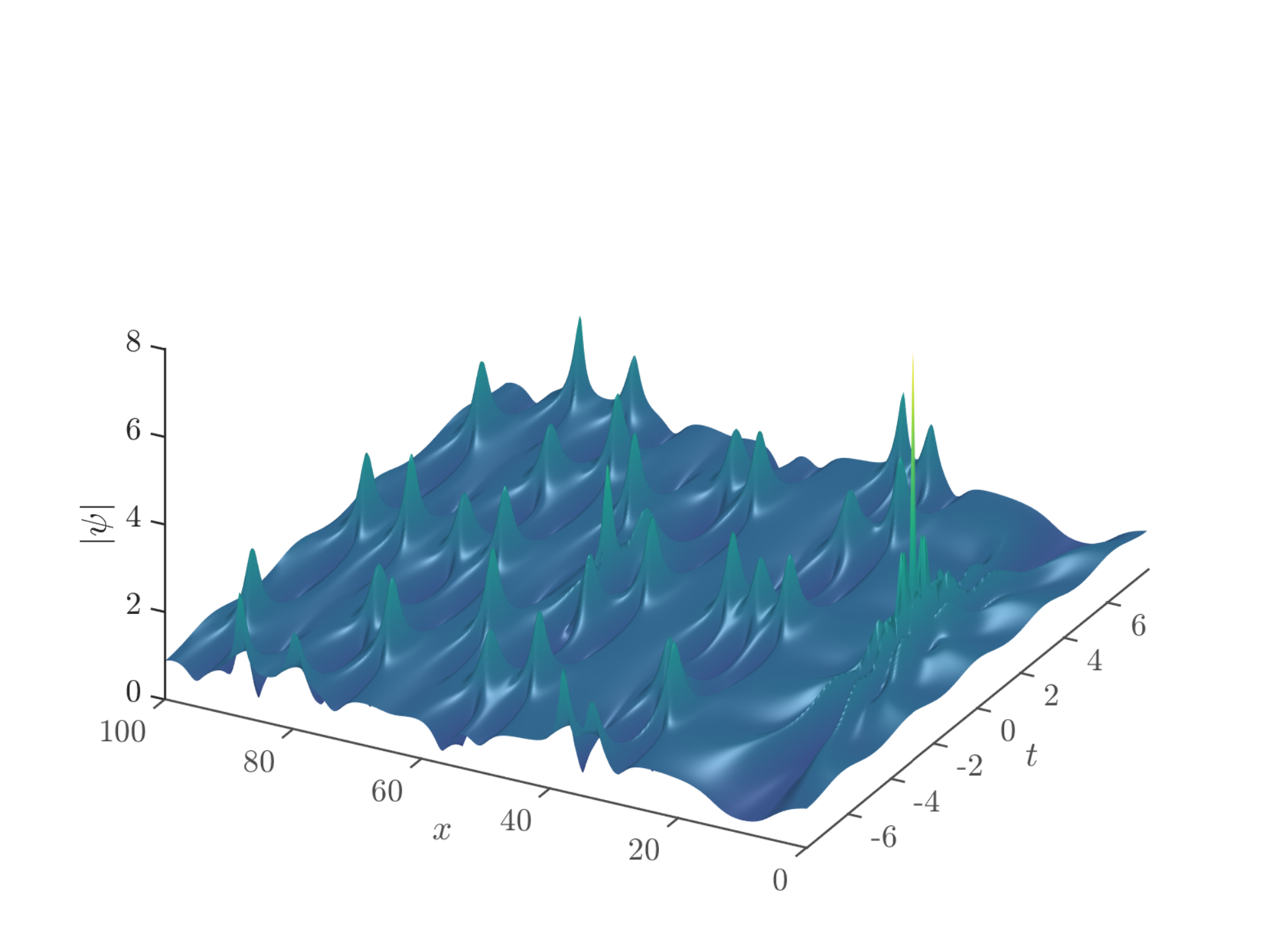}
    \caption{The result of Example \ref{lst:ex8}. One can see the fifth-order Akhmediev breather formed at $x$ = 10 as expected, followed by smaller peaks due to modulation instability.}
    \label{fig:DT_0}
\end{figure}

In this example, we combine the Darboux transformation with simulations to ``dynamically'' generate a higher-order solution of the NLSE, as done in Ref. \cite{Chin2015, Nikolic2017, Nikolic2019} and others. We offer the utility function \jlinl{ψ₀\_DT} to generate these initial conditions from whichever desired \emph{breather on a uniform background}. Using this function with other seeds is not currently supported but can be done manually with ease, as long as care is taken with the boundary conditions. The full example is shown in code listing \ref{lst:ex8} and is quite similar to the previous examples. The result is depicted in Fig. \ref{fig:DT_0}.

\begin{lstlisting}[language=Julia, caption={Solving the NLSE with an initial condition from the Darboux Transformation}, captionpos=b, label={lst:ex8}]
λ₁ = 0.98im
λ, T, Ω = params(λ = λ₁)

xᵣ = 0=>100
box = Box(xᵣ, T, dx=1e-3, Nₜ = 512, n_periods = 1)

λ = λ_maximal(λ₁, 5) # array of 5 eigenvalues
xₛ = [0.0, 0.0, 0.0, 0.0, 0.0]
tₛ = [0.0, 0.0, 0.0, 0.0, 0.0]
ψ₀ = ψ₀_DT(λ, tₛ, xₛ, -10, box) #Extract ψ from the DT at x=-10

sim = Sim(λ₁, box, ψ₀, T4A_TJ!)

solve!(sim)
compute_IoM!(sim)
\end{lstlisting}

\subsubsection{Example 9: Breather to Soliton Conversion in a Fifth-Order NLSE \label{sec:ex9}}

Given a generalized fifth-order NLSE of the form \eqref{eq:extended_NLSE}, we can compute the Darboux transformation by passing a dictionary \texttt{f} to the \texttt{Calc} constructor as follows

\begin{lstlisting}
f = Dict{Symbol,Float64}(:α=> 0.75, :γ => -0.12, :δ=>-0.13)
calc = Calc(λ, tₛ, xₛ, seed, box, f = f) 
\end{lstlisting}

Writing the eigenvalue as $\lambda = v + i \nu$, one can impose some constraints on the real part $v$ as a function of $\alpha$, $\gamma$, $\delta$ and $\nu$ to convert a breather to a soliton \cite{Chowdury2015, Nikolic2019}. Note that this is not possible in the simple case of the cubic NLSE. This constraint takes the form

\begin{align}
    64 \delta v^3 -24 \gamma v^2 - 8(\alpha + 2\delta + 8\delta \nu^2)v + 4\gamma(1 + 2 \nu^2) + 1 = 0.
\end{align}

This is a simple cubic polynomial, and the real root gives the value of $v$, which guarantees breather to soliton conversion. This functionality is implemented in \texttt{NonlinearSchrodinger.jl} via the function $\lambda$\texttt{\_given\_f} which can be used as shown

\begin{lstlisting}
f = Dict{Symbol,Float64}(:α=> 0.75, :γ => -0.12, :δ=>-0.13)
ν = 0.9
λ = [λ_given_f(f, ν)]
\end{lstlisting}

\begin{figure}
    \centering
    \includegraphics[width=0.97\linewidth]{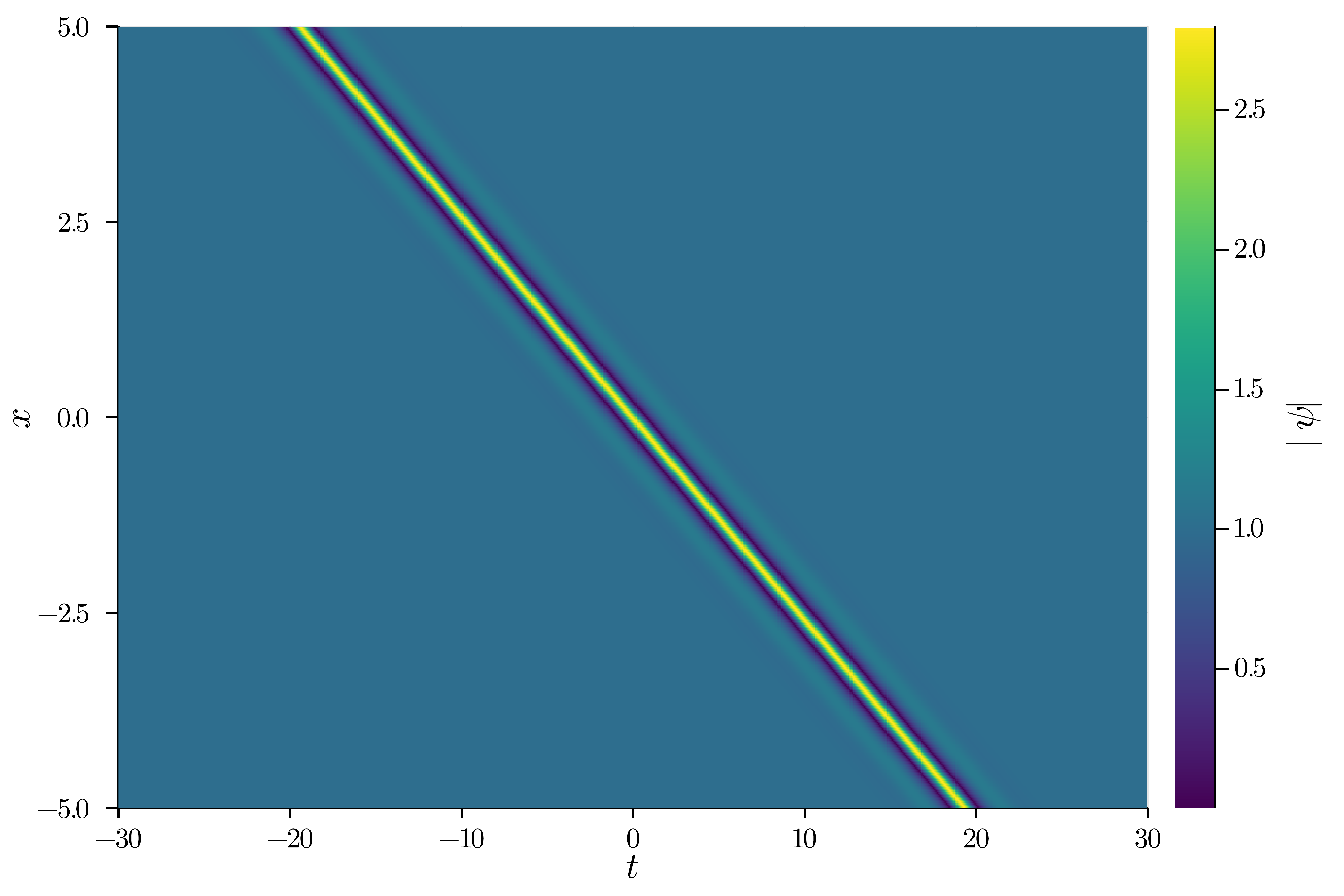}
    \caption{Breather to soliton conversion in a fifth-order extended NLSE as computed in Example \ref{lst:ex9}.}
    \label{fig:ex9}
\end{figure}

This function uses Skowron and Gould's algorithm \cite{Skowron2012}, as implemented in Julia's \texttt{PolynomialRoots.jl} package. The rest of the computation is similar to any other Darboux transformation example and is shown in code listing \ref{lst:ex9}. We stress that this example uses the breather seed of Sec. \ref{sec:seed_exp}, not the soliton seed of Sec. \ref{sec:seed_0}, yet the resulting solution is a soliton, not a breather. This conversion is a fascinating feature of these extended NLSEs. The resulting soliton is depicted in Fig. \ref{fig:ex9}.

\begin{lstlisting}[language=Julia, caption={Breather to soliton conversion in a fifth-order extended NLSE.}, captionpos=b, float, floatplacement=H!, label={lst:ex9}]
xᵣ = -5=>5
T = 60
box = Box(xᵣ, T, Nₓ=1000, Nₜ = 1000)

f = Dict{Symbol,Float64}(:α=> 0.75, :γ => -0.12, :δ=>-0.13)
λ = [λ_given_f(f, 0.9)]
xₛ = [0.0]
tₛ = [0.0]

seed = "exp"
calc = Calc(λ, tₛ, xₛ, seed, box, f = f) 

solve!(calc)
\end{lstlisting}

\section{Future Directions and Conclusion \label{sec:conc}}

There are many possible enhancements planned for \texttt{NonlinearSchrodinger.jl}'s future. First of all, we plan to implement several memory optimizations to aid in the performance of long-``time'' simulations with a fine step $dx$. Moreover, we plan on supporting the numerical integration of more nonlinear Schr\"odinger-type equations such as the Hirota equation and the Sasa-Satsuma equation. 

While our current implementation of the Darboux Transformation supports extended nonlinear Schr\"odinger equations of order up to 5, we plan to implement the entire hierarchy \cite{Kedziora2015} so that calculations can be performed for arbitrary order equations. Furthermore, we plan to add support for cnoidal seeds beyond the cubic NLSE.

Finally, we plan to support quadruple-precision floating-point numbers to exploit higher-order integrators fully. We have already shown in Sec. \ref{sec:benchmark} that eighth order algorithms are bottlenecked by double-precision. Quadruple-precision will alleviate this issue and allow for the implementation of even higher-order integrators. 

Additionally, quadruple-precision will enable the calculation of much higher order solutions via the Darboux transformation. It is well known that double-precision can only handle breather and soliton solutions via the Darboux transformation up to order $N \sim 30$ \cite{Gelash2018}. Quadruple-precision calculations will make the package more versatile and allow for computing, e.g., breather and soliton gas solutions with $N \sim 100$ \cite{Gelash2018,Roberti2021}.

In conclusion, we have presented a powerful and unique package that allows one to study numerical and analytical solutions of nonlinear Schr\"odinger equations via higher-order integrators and Darboux transformations. The package provides a simple interface and makes it straightforward to study complicated solutions and use optimal symplectic and RKN integrators up to eighth order. Moreover, we offer many utilities for studying maximal intensity families, nonlinear Talbot carpets, breather to soliton conversion, and visualization of the solutions.

\section*{Acknowledgements}
We are indebted to Siu A. Chin for the in-depth conversations on this manuscript, higher-order integrators, and much else. We thank Tomohiro Soejima for the insightful discussions about code optimization in Julia. We acknowledge the helpful talks with Milivoj R. Beli\'c and Stanko N. Nikoli\'c. We are grateful to Steven G. Louie for supporting this project.

\paragraph{Funding information}
O.A.A is supported by the UC Berkeley Physics Department.

\begin{appendix}
\section{Installation Instructions \label{app:install}}
After Julia is properly installed, it is quite simple to install \texttt{NonlinearSchrodinger.jl} as it is listed in Julia's general repository. Julia comes with its own package manager and, starting from a terminal session; the package can be installed as follows:
\begin{lstlisting}
$ julia
julia> ]
pkg> add NonlinearSchrodinger
\end{lstlisting}

The first command is run from the terminal, and the second command instantiates Julia's built-in package manager. The third command installs the package from the general repository. No additional libraries or any other software are needed.

The following command must be issued once per session to use the package. 

\begin{lstlisting}
julia> using NonlinearSchrodinger
\end{lstlisting}

It is assumed in all exercises that it has already been run.

The \texttt{Plots.jl} package, used for plotting as outlined in Sec. \ref{sec:ex1}, is installed and used in the same way.

To input Greek letters (Unicode characters) in Julia, such as those shown in the exercises, enter them in the same way you would in \LaTeX\, followed by the tab key. For example, to type $\lambda$ in a Julia session, type the following: \texttt{\textbackslash lambda<TAB>}. Please consult the Julia documentation for more details.

\end{appendix}

\bibliography{main.bib}

\nolinenumbers

\end{document}